\documentclass[10pt,journal,compsoc]{IEEEtran}
\usepackage{orcidlink}
\usepackage[para]{footmisc}
\usepackage{ragged2e}
\usepackage{etoolbox}
\patchcmd\makefootnoteparagraph{\par}{\RaggedRight\par}{}{}

\usepackage{graphicx}
\usepackage{graphics}

\usepackage[ngerman, english]{babel}
\usepackage{todonotes, soul}

\newcommand{\para}[1]{\vspace{1mm}\noindent\textbf{#1.}}

\usepackage{colortbl}
\usepackage{xcolor}
\definecolor{mygreen}{RGB}{5, 114, 63}
\definecolor{myblue}{RGB}{57, 128, 198}
\definecolor{myred}{RGB}{201, 4, 30}

\usepackage{amsthm}
\usepackage{amsmath} 
\usepackage{amssymb}
\usepackage{lipsum,mathtools}
\usepackage{relsize}
 \usepackage[export]{adjustbox}
 \usepackage{enumitem}

\usepackage{multirow}
 \usepackage{booktabs}
\usepackage{hhline}

\usepackage{subcaption}
\newcounter{ChristosNOC}
\stepcounter{ChristosNOC}

\newcounter{RihanNOC}
\stepcounter{RihanNOC}

\DeclareUnicodeCharacter{0303}{\textdegree}

\newcounter{JarkeNOC}
\stepcounter{JarkeNOC}

 \newcommand{\new}[1]{\textcolor{black}{#1}}

\ifCLASSOPTIONcompsoc
  \usepackage[nocompress]{cite}
\else
  \usepackage{cite}
\fi
\ifCLASSINFOpdf
\else
\fi

\begin{document}
%
\title{Data Lakes: A Survey of Functions and Systems}
%
%
%
%

\author{Rihan~Hai~$^{\orcidlink{0000-0002-3720-6585}}$,~Christos~Koutras~$^{\orcidlink{0000-0003-3015-154X}}$,~Christoph~Quix~$^{\orcidlink{0000-0002-1698-4345}}$,~and~Matthias~Jarke~$^{\orcidlink{0000-0001-6169-2942}}$,~\IEEEmembership{Lifetime Senior Member, IEEE}
\IEEEcompsocitemizethanks{\IEEEcompsocthanksitem R. Hai and C. Koutras are with the Department of Software Technology, Delft University of Technology, Delft, Netherlands.\\
E-mail:\{r.hai, c.koutras\}@tudelft.nl.


\IEEEcompsocthanksitem C. Quix is with Hochschule Niederrhein, Krefeld, Germany and Fraunhofer FIT, Sankt Augustin, Germany.\\
E-mail:christoph.quix@hs-niederrhein.de.
\IEEEcompsocthanksitem M. Jarke is with RWTH Aachen University, Aachen, Germany and Fraunhofer FIT, Sankt Augustin, Germany.\\
E-mail:jarke@dbis.rwth-aachen.de.}
}

%
%

\markboth{Journal of \LaTeX\ Class Files,~Vol.~14, No.~8, August~2015}%
{Shell \MakeLowercase{\textit{et al.}}: Bare Demo of IEEEtran.cls for Computer Society Journals}
%



\IEEEtitleabstractindextext{%
\begin{abstract}
Data lakes  are becoming increasingly prevalent for big data management and data analytics.
In contrast to traditional `schema-on-write' approaches such as data warehouses,
data lakes are repositories  storing  raw data in its original formats and providing  a common access interface.  
Despite the strong interest raised from both academia and industry, there is a large body of ambiguity regarding the definition, functions and available technologies for data lakes. A complete, coherent  picture of
data lake challenges and solutions is still missing.
This survey reviews the development, architectures, and systems of data lakes. 
We provide a comprehensive overview of research questions for designing and building data lakes. 
We classify the existing approaches and systems  based on their provided functions for data lakes, which makes this survey a useful technical reference for designing, implementing and deploying data lakes. 
We hope that the thorough comparison of existing  solutions and the discussion of open research challenges in this survey will motivate the future development of data lake research and practice. 
\end{abstract}

}

\maketitle
  \section{Introduction}
\label{sec:background}


\IEEEPARstart{B}{ig} data has undoubtedly become one of the most important challenges in database
research.
Unprecedented volume, large variety, and high velocity of data impede their collection, storage, and processing;  
especially the variety of data still poses a 
daunting challenge with many open issues \cite{DBLP:journals/cacm/AbadiAABBCCDDFG16}.
Web-based business transactions, sensor networks, real-time
streaming, social media, and scientific research generate a large amount of
(semi-)structured and unstructured data, often stored in separate information silos. 
Combining and integrating the information across
these silos is critical for reaching valuable insights.

%

Traditional  \emph{schema-on-write}
approaches, like
the extract, transform, load (ETL) process of data warehouses \cite{JLVV03},  are  inefficient for such 
data management requirements. This has drawn the interest of many developers and
researchers to NoSQL data management systems. 
NoSQL  systems   provide
data management features tailored to high amounts of schema-less data, which enables a 
\emph{schema-on-read}
manner of data handling, i.e., the structure of data is not required for storing but only when further analyzing and processing the data. 
Open-source platforms, such as Hadoop \cite{shvachko2010hadoop}
 with higher-level languages (e.g., 
Pig and Hive), as well as NoSQL databases (e.g.,  
MongoDB and Neo4j), have gained popularity. 
Although the current market share is still dominated by relational database systems,
a one-size-fits-all big data system is unlikely to solve all the challenges related to data management today. 

To address this gap, \emph{data lakes} (DLs) have been proposed. 
In essence, a data lake is a flexible, scalable data storage and management system,  
	which ingests and stores raw data from heterogeneous sources in their original format, 
	and provides maintenance, query processing and data analytics 
	in an on-the-fly manner, with the help of rich metadata~\cite{DBLP:conf/bdcloud/WalkerA15, pwcdatalake2015, terrizzanodata, quix2016gemms}.
Data lakes are proposed to store and manage data in many real-life use cases: Internet of things (IoT) and smart city \cite{DBLP:conf/icde/MehmoodGCKBVTR19}, 
manufacturing \cite{DBLP:conf/icphys/PennekampGHMQHG19},
medicine \cite{DBLP:conf/dils/GeislerQHA17, DBLP:conf/semweb/QuixGHA17, eder2021data}, 
 mobility   service (e.g., Uber) \cite{10.1145/3448016.3457552}, 
biology \cite{10.3389/fbioe.2020.553904}, 
smart grids \cite{DBLP:journals/access/MunshiM18, 9372181}, air quality control \cite{DBLP:conf/dmbd/WibowoSS17}, flights data \cite{martinez2017integrating}, disease control, labor markets and products \cite{BociBigDataArchitecture}.

\subsection{Survey goal and related work}

In the past decade, various solutions and systems have been proposed to address the research challenges of data lakes.
However, while `data lake' is a current buzzword with a lot of hype surrounding it, there is a lot of ambiguity about its exact definition and functions.
Moreover, most recent data lake proposals only target a specific research problem or certain types of source data. A coherent, complete picture
of data lake problems and solutions is still missing.

In this  survey, we provide  a thorough explanation of the data lake concept, its development, more importantly, a categorization and review of existing data lake solutions. The survey also aims at helping researchers and developers to build or customize a data lake, and discover open questions and future research directions about data lakes.
\new{Earlier efforts in structuring the data lake field 
only provide a limited view on a subset of research problems regarding data lakes. Moreover, none of these works touch on the details of future data lake challenges such as supporting machine learning in data lakes.} 

Several earlier works \cite{DBLP:journals/dbsk/Mathis17, DBLP:conf/birthday/JarkeQ17,  DBLP:reference/bdt/QuixH19} propose   possible research topics that should be included in data lakes without reviewing  existing systems for data lake. These works are orthogonal to our goals in this survey. Instead of merely listing potential research questions, our focus is to make a technical comparison of the existing systems for data lakes. 

In a recent tutorial, Nargesian et al. \cite{nargesian12data} cover seven  functions of data lakes. \new{They have briefly discussed   existing data lake solutions, together with technologies and systems potentially useful for data lakes\footnote{\url{https://rjmillerlab.github.io/data-lake-tutorial-slides/}}. 
In contrast to this tutorial, our survey provides a more holistic introduction to data lakes and discusses the required functions of data lakes in more detail (cf. Fig.~\ref{fig:rw-high-view} and Table~\ref{tbl:all} for our functional view of data lakes). 
}


Couto et al. \cite{couto2019mapping} compare different data lake definitions, and list common open-source tools used in data lake architectures. In \cite{zagan2020data}, Zagan et al. review the architectures of seven specific data lake systems. In a data lake architecture proposal \cite{10.1007/978-3-030-27615-7_23}, Ravat and Zhao propose classification criteria for metadata categories and  data governance in data lakes. Giebler et al. \cite{Giebler2019} discuss some additional aspects of data lake architecture, data storage, data modeling, metadata management and data governance. 
Each of these works only provides a partial list of data lake functions.
In this survey, we  give a more comprehensive view of the current data lake landscape, and   have a more in-depth discussion  regarding research challenges and solutions of data lakes.  

In \cite{sawadogo2020data} Sawadogo et al. compare different data lake definitions, architectures, metadata types, metadata models, and metadata management components (e.g., semantic enrichment). They provide a high-level guide for conceptual design of data lakes. However, their discussion regarding functions to implement for a data lake system is very brief and limited to summarizing the  
 open-source technologies and tools used in existing data lakes,  e.g., Apache Spark, Drill, and Pig. A similar survey \cite{cherradi2021data} also only focuses on  the aspects of data lake architecture, metadata management, and open-source technologies.
In this survey, we cover a wider range of topics on data lakes \emph{beyond} architecture and metadata management. We also show how to navigate from a conceptual architecture to system functions. 
We propose a more fine-grained categorization of existing   systems for data lakes, and provide a more detailed comparison of systems in each category. 
For each data lake function, we also cover the state-of-the-art  systems not mentioned in  \cite{sawadogo2020data, cherradi2021data}.

\subsection{Contributions and outline}



Our main contributions  are summarized as follows: 
\begin{itemize}[leftmargin=*]
	\item 
	 We review the more than ten-year development of the data lake concept and implementations, \new{and discuss future directions.} 
	\item 
\new{We clarify the workflow and functions for building a data lake through a  fine-grained architecture.}
	
	 \item  We provide a three-level classification of existing studies about data lakes according to their provided functions. 
     We analyze each class of research problems in depth, and compare the existing data lake approaches. 
\end{itemize}
 
  \begin{figure}[tb]
	\begin{center}
		\includegraphics[width=\columnwidth]{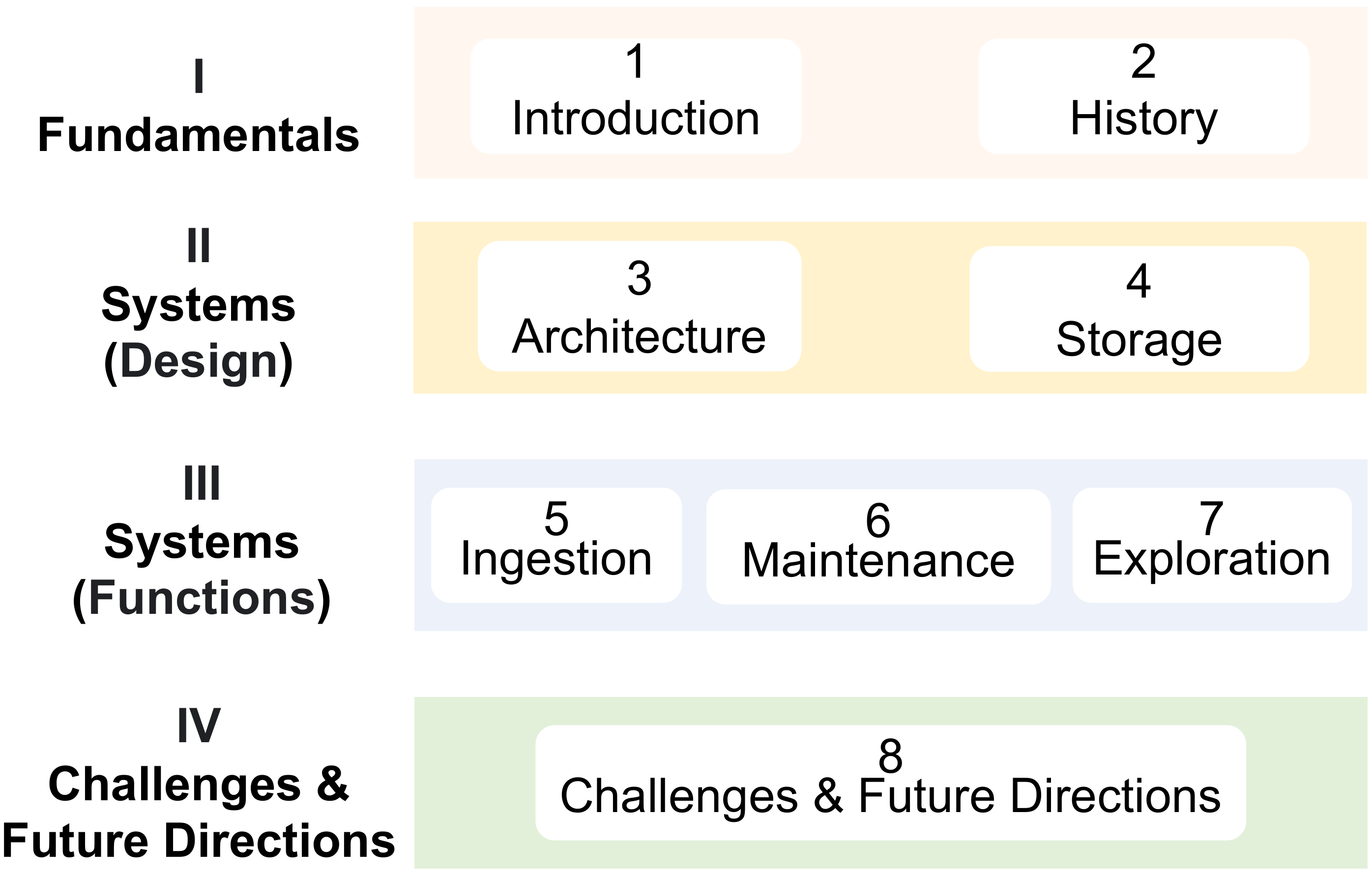}
			\vspace{-0.2cm}	
		\caption{Survey outline}
		\label{fig:outline}
		\vspace{-0.8cm}
	\end{center}
	
\end{figure}
\para{Scope of the survey} In this survey, we focus on systems that explicitly claim to be a data lake (e.g., personal data lake \cite{DBLP:conf/bdcloud/WalkerA15}), 
or  provide partial functions of a data lake (e.g., the data discovery system Aurum \cite{DBLP:conf/icde/FernandezAKYMS18}). 
It is beyond the scope of \textbf{one} survey article to list all \emph{possible} research topics and all \emph{potential} solutions for data lake systems.
Some research topics mentioned in this survey have been intensively studied in the database community, e.g., data integration, data cleaning, and data discovery. 
There are dedicated surveys on these topics, while in this survey we only introduce and compare systems tackling these problems within a data lake.
For these topics, we will explain 
the lake-specific problem settings. For instance, consider the data integration problem (Sec.~\ref{ssec:di}), which is to resolve heterogeneous schemata or entity values.
In a data lake, for data integration we often assume more levels of heterogeneity among the data sources.
Finally, for system comparison, we mainly cover \textbf{implemented} systems  that resolve  \textbf{research} problems of data lakes, rather than high-level  DL  system proposals or commercial DL products. We made this choice because such high-level proposals often lack details for meaningful comparison, and do not always reflect the feasibility of actual system implementation. For a more industrial point of view, we point the reader to papers like \cite{russom2017data, gorelik2019enterprise, hukkeri2020study}. 

 

\para{Outline}
As illustrated in Fig.~\ref{fig:outline},  the survey has four parts. 
The first part covers the fundamentals of data lakes, including the introduction 
(Sec.~\ref{sec:background}) and the origin and development  (Sec.~\ref{sec:history}). 
Then we discuss the common aspects that almost  every data lake designer  needs to consider: the system architecture (Sec.~\ref{sec:archit}) and data storage (Sec.~\ref{ssec:DLS}). 
In particular, we introduce the criteria of classifying data lake  solutions  in Sec.~\ref{ssec:3l_criteria}. 
In the third part,  we categorize existing  data lake solutions by their functional tiers: ingestion (Sec.~\ref{sec:ing}), maintenance (Sec.~\ref{ssec:maint}), and exploration (Sec.~\ref{sec:exp}). 
Finally, we discuss research challenges and future directions in Sec.~\ref{sec:chall} and 
conclude the survey in Sec.~\ref{sec:con}.

\para{How to use this survey} We organize the survey in the structure of Fig.~\ref{fig:outline}, such that it is self-contained and presented in a natural flow. We first explain     high-level concepts and architecture, before discussing   data storage options and  functions.
A data lake expert interested in a particular research problem, can directly go to Sec.~\ref{sec:ing}-\ref{sec:exp}. 
Discussions on challenging new directions are in Sec.~\ref{sec:chall}.

\section{A brief history of data lakes}
\label{sec:history}
As of this writing, the concept of data lakes is about a decade old and has significantly evolved in this period. We summarize this evolution in three stages.


\subsection{2010–2013: Beginnings}
The concept of \emph{data lake} was first coined in the industry.
In 2010,  it was first proposed by  \emph{Pentaho} CTO James Dixon, as a solution
that handles raw data from \textbf{one} source and supports diverse user requirements \cite{Dixon}. 
This was seen in sharp contrast to data warehouses or data marts for which the structure and usage of the data must be predefined and fixed, and rigorous data extraction, transformation, and cleaning are necessary before entering data. By storing \textbf{raw} data in the  original format, 
data lakes could avoid or delay this expensive standard preprocessing.

In 2013, \emph{Pivot}  proposed an architecture for a business data lake \cite{Pivotal2013},  
which ingests \textbf{multiple} data sources in three abstract tiers:
(1) an \emph{ingestion tier} takes data in real-time/micro-batch/batch,  
(2) an \emph{insight  tier}  analyzes data in real-time or interactive time and derives insights, and (3) an \emph{action tier}  that links  insights with the existing applications;
additional tiers \emph{monitor and manage} the data.
Pivot also suggested using Hadoop \cite{shvachko2010hadoop} as the storage system of a data lake, 
and applying its existing products to realize  the previous tiers.
However, not many details were given w.r.t. the actual implementation of a data lake. 

\subsection{2014–2015: Criticisms and further development}
In 2014, \emph{Gartner} raised several criticisms about data lakes \cite{gartner2014}.  
The main one was that ingesting disparate data might easily turn the data lake into an unusable ``data swamp'', unless there are metadata management and data governance. In particular, after ingestion, the semantics and data quality of the raw data are unknown, and the origin (provenance) of individual datasets and its possible connection  to other datasets are missing. Indeed missing this information hinders user interaction with the data lake. 
In addition, Gartner pointed out that the existing data lake solutions did not provide a good answer  
on how oversight of data security and privacy should be conducted.  
These crucial criticisms had a significant influence on  many data lakes studies in the following years, 
which we discuss in Sec.~\ref{sec:ing}-\ref{ssec:maint}. In response, Dixon revisited the general concept \cite{Dixon2014}, and emphasized that a data lake should also be equipped with metadata and governance, so that even with data in its raw form, a data lake could enable ad-hoc data analytics.

As more DL proposals started to emerge, they brought new requirements, solutions, and challenges. \new{They significantly augmented the \emph{possible} functions of a data lake, e.g.,  heterogeneous data, schema-on-read, metadata extraction/enrichment/management, applied Artificial Intelligence, and Crowdsourcing.}
\emph{PwC} defined a data lake as a repository of structured, semi-structured, and unstructured data in heterogeneous formats \cite{pwcdatalake2015}, originating from the business transactions, sensors, or mobile/cloud-based applications. 
With Hadoop  in the center,  \new{a new requirement is that} a data lake should provide a low-cost data storage that is easy to access, yet in a \textbf{schema-on-read} manner,
i.e., the data and metadata (e.g., semantics) 
can grow over time. 
They postulate that a data lake actively extracts metadata from the raw data and stores it; 
then, it discovers 
patterns in the raw data. 
Moreover, users provide additional descriptive information of datasets 
(e.g., semantic annotations, domain-specific knowledge, and attribute linkages). 
The dynamic interaction between the data lake and users should thus continuously improve the quality and value of data.

Other proposals addressed new possibilities such as Artificial Intelligence (AI) and Crowdsourcing to facilitate  data integration, access,  and quality improvement in data lakes \cite{DBLP:journals/expert/OLeary14}. 
For example, AI helps with extracting features of data, generating tags with descriptive metadata, 
finding related datasets, discovering possible structures from schema-less data, and avoiding data redundancy. 
Crowdsourcing can help with collectively tagging semantic knowledge about the data, and
linking possible relationships among datasets.
With a special focus on security information and event management, 
In \cite{Marty2015} Marty  discussed how to properly store and access the data. 
The importance of \emph{metadata management} is emphasized in \cite{DBLP:conf/bdcloud/WalkerA15} 
with an architecture to  parse, store, and query  diversely structured personal data. 
Another proposal  \cite{fang2015managing} emphasized the importance of Human-in-the-loop, e.g., data scientists govern the data in  data lakes. 

\subsection{2016–present: Prosperity and diversity}
Since 2016 the realization of data lakes in industry and research has been booming. 
There are proposals about data lake architectures \cite{architectingDL2016, inmon2016data, DBLP:conf/medes/MaderaL16, Singh2019}, 
 concept, components and challenges \cite{miloslavskaya2016big, DBLP:journals/dbsk/Mathis17, DBLP:reference/bdt/QuixH19}. 
%

Many IT companies offer commercial tools for building data lakes, e.g., 
\emph{GOODS} \cite{DBLP:journals/debu/HalevyKNOPRW16} from Google,
 \emph{Azure Data Lake} \cite{DBLP:conf/sigmod/RamakrishnanSDK17} from Microsoft, \emph{AWS Lake}\footnote{\url{https://aws.amazon.com/lake-formation/}} from AMAZON,   \emph{Vora} from SAP \cite{DBLP:conf/btw/SengstockM17}, 
 IBM and Cloudera\footnote{\url{https://www.ibm.com/analytics/data-lake}}, 
 Oracle\footnote{\url{https://www.oracle.com/data-lakehouse/}}, 
and \emph{Snowflake}\footnote{\url{https://www.snowflake.com/data-lake/}}.
%
\emph{Delta Lake}\footnote{\url{https://delta.io/}} from Databricks is an open-source project that 
  offers a storage tier compatible with  Spark\footnote{\url{https://spark.apache.org/}} APIs. 

Meanwhile, data lake related research problems are raising massive attention associated with the implementation of DL prototypes.
A large range of challenges are discussed such as metadata management \cite{hai2016constance},
data quality \cite{farid2016clams}, data provenance \cite{DBLP:conf/eScience/SuriarachchiP16}, metadata enrichment \cite{DBLP:conf/cikm/BeheshtiBNCXZ17, DBLP:conf/er/HaiQW19}, 
dataset organization \cite{DBLP:conf/sisap/AlserafiCA017, DBLP:conf/sigmod/ZhuDNM19}, 
modeling \cite{quix2016metadata, DBLP:conf/ideas/NogueiraRD18, DBLP:conf/er/Giebler19}, data integration \cite{DBLP:conf/adbis/HaiQZ18, hai2019rewrite} and related dataset discovery \cite{DBLP:conf/icde/FernandezAKYMS18, DBLP:conf/sigmod/BrackenburyLMEU18, DBLP:conf/sigmod/ZhuDNM19, bogatu2020dataset, zhang2020finding}. 
Such systems 
targeting  specific research challenges of data lakes, are also our main focus in this survey. We  address them in   Sec.~\ref{ssec:DLS}-\ref{sec:chall}.

 \section{Data lake architecture and proposed categorization criteria}
\label{sec:archit}
The \emph{architecture} of a data lake describes the structure and components of the system, 
indicating how to store, organize and use the data. 
A recent survey \cite{sawadogo2020data} elaborated on the categorization of existing data lake architectures, while a methodology for designing data lake architecture is discussed in \cite{Giebler2021}. 
In existing data lake architecture proposals, there are mainly two high-level data lake  philosophies, \emph{pond} and \emph{zone} architectures. 
Rather than repeating similar content as in \cite{sawadogo2020data, Giebler2021}, 
we briefly review pond and zone architectures in Sec.~\ref{ssec:pond_zone}. 
We mainly focus on   presenting an integrative, function-oriented architecture and classification criteria for data lake studies in Sec.~\ref{ssec:3l_criteria}. 
Additionally, we discuss the different kinds of data lake users in Sec.~\ref{sec:users}.

\subsection{Pond and zone architectures}
\label{ssec:pond_zone}

The \emph{pond  architecture} \cite{inmon2016data} partitions  ingested data  by their status and usage. In specific, ingested data is first stored in the \emph{raw data pond}, then transformed and moved to 
the \emph{analog data pond}, \emph{application data pond}, or \emph{textual data pond} if possible. 
Associated processes are created to prepare the data for future analytical processing. Later on, valuable data is secured long-term in an \emph{archival data pond}.
For instance, analog data generated by an automated device is moved to the analog data pond followed by data reduction to a feasible data volume.
In contrast, the \emph{zone architecture} \cite{Chessell2014, 2014big, architectingDL2016, Patel2017, 10.1007/978-3-030-27615-7_23}, 
separates the life cycle of each dataset into different stages. For instance, there could be individual zones for loading data and checking data quality, storing raw data, storing cleaned and validated data, 
discovering and exploring the data, or using the data for business/research analysis. 


\begin{figure*}[tb]
	\begin{center}
		\includegraphics[width=0.84\linewidth]{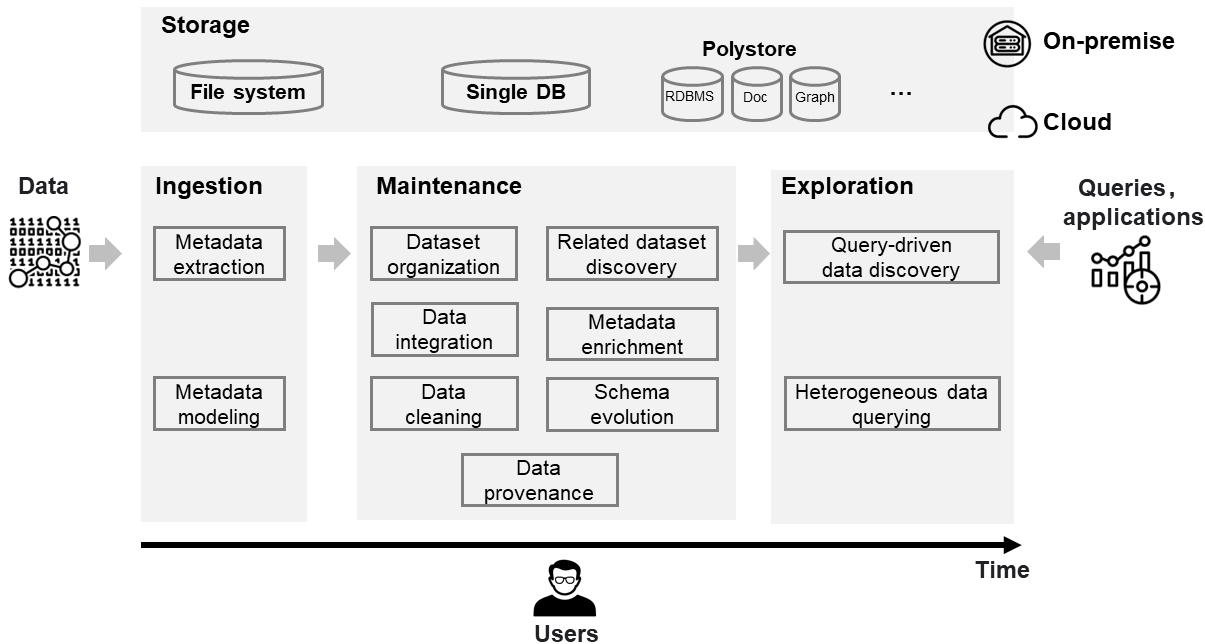}
		\vspace{-0.2cm}
		\caption{Proposed architecture for data lake solution categorization}
		\label{fig:rw-high-view}
	\vspace{-0.5cm}
	\end{center}
	
\end{figure*}

\subsection{Proposed architecture and  categorization criteria}
\label{ssec:3l_criteria} 
\label{ssec:3l_Archit} 
High-level architectural philosophies, such as pond or zone architecture, often lack technical details about \textbf{functions}, which hampers modular and reusable implementations. 
Therefore, we propose  an architecture based on our previous works \cite{hai2016constance, DBLP:conf/birthday/JarkeQ17}, as  in Fig.~\ref{fig:rw-high-view}. 
This architecture can also be seen as an abstraction of earlier tier-based data lake proposals \cite{DBLP:conf/eScience/SuriarachchiP16,architectingDL2016,  DBLP:conf/dexa/EndrisRVA19}.
Notably, here we define specific functions in each tier, which are not covered in these proposals. 
The goal of proposing such an architecture is two-fold. First, it clarifies  the  necessary functions in the whole workflow of a data lake, and provides a more comprehensive  view compared to earlier DL architecture proposals. 
Second, it enables a fine-grained categorization and in-depth comparison of    
existing systems for each function.

\para{Proposed architecture} 
The data lake architecture in Fig.~\ref{fig:rw-high-view} provides a three-tier function-oriented linkage between  
data lake users, and a storage tier encompassing potentially multiple different technologies. 
In Sec.~\ref{ssec:DLS} we review storage strategies applied in existing data lake solutions, which could be on-premise or cloud, single  or multiple data storage systems, relational  or NoSQL databases.

The architecture divides   functions of the whole workflow into three tiers, according  to \emph{when} the functions are needed. 
There are certain functions during or right after data is ingested, which are in the \emph{ingestion tier}. 
The  ingestion  tier is responsible for importing data from heterogeneous sources
into the data lake system. The main challenges are about extracting and modeling metadata (Sec.~\ref{sec:ing}). 

Functions in the middle (i.e., the \emph{maintenance tier}) are for  general management and organization of  ingested datasets, which can also be  considered as the preparation for querying.
To prepare ingested raw data for querying or analytics, a data lake needs a set of operations in the maintenance tier to organize, discover, or integrate   datasets (Sec.~\ref{ssec:maint}). 

Some functions are triggered by user queries or applications of a data lake,  as  shown on the right side of Fig.~\ref{fig:rw-high-view}, i.e., the \emph{exploration tier}. 
These functions mainly contribute to allowing users to access the data lake. We observe  two manners of exploring in existing works: query-driven data discovery and  heterogeneous data querying, which we discuss in Sec.~\ref{sec:exp}. They also form the basis for external application tiers on top of these three tiers, e.g., for visualization \cite{DBLP:conf/icde/MehmoodGCKBVTR19} or machine learning. 

\para{Three-level classification criteria}
As one of our main contributions, here we explain our classification criteria applied in this survey, which is based on our proposed data lake architecture.
Many existing  research works focus on one specific function inside a data lake. 
Thus, we classify them by their functions in the architecture in   Fig.~\ref{fig:rw-high-view}. 
Table~\ref{tbl:all} shows such a categorization:
we first group the existing systems by the tier and then the exact provided function. 
Of course, some  systems provide more than one function. For instance, some data discovery systems in Table~\ref{tbl:all} also have components for metadata extraction and querying. 
We will give a detailed explanation of each functional criterion in Sec.~\ref{sec:ing}-\ref{sec:exp}. 
For some extensively studied functions, e.g., metadata modeling, related dataset discovery, we further categorize the systems based on their methods, which will be explained in the corresponding subsections. 
Notably, as emphasized in Sec.~\ref{sec:background}, in this survey we mainly compare existing systems developed for data lakes. The 11 functions in Table~\ref{tbl:all} are summarized based on existing data lake related studies. For more potentially useful functions in a data lake architecture, yet not studied, see \cite{DBLP:reference/bdt/QuixH19}.
%
In summary, we follow a three-level categorization of existing data lake solutions: tiers (\emph{when} the function is needed), functions (\emph{what} the function is), and methods (\emph{how} the function is achieved).

%

\para{Lake-specific research perspectives} 
Some functions in Table~\ref{tbl:all} are challenges almost every  data management system with heterogeneous data needs to face, e.g., related dataset discovery, metadata extraction and enrichment,  data cleaning, and data provenance. However, as discussed in Sec.~\ref{sec:background}, data lakes require a novel, flexible manner of data management, which also leads to new research challenges. By ingesting the raw data as it is, we cannot simply refer to an existing schema. Instead, we need to actively discover related datasets partially at ingestion time, but mostly during maintenance or even at querying time. 
Therefore, we discuss related  dataset discovery methods from both maintenance (Sec.~\ref{sssec:relDS}) and exploration (Sec.~\ref{ssec:qds}) perspectives. 
Moreover, a data lake often needs to ingest a large volume of data,  possibly also at a high velocity or even as continuous data streams, which cannot be stored in full in the data lake 
Not all metadata can be extracted at ingestion time (Sec.~\ref{sec:meta_extr}), but we need to continue enrichment during later phases as well (Sec.~\ref{sec:metad_enrich}). For similar reasons, we assign the continuous activities of data cleaning (Sec.~\ref{ssec:dq_improv}) and data provenance (Sec.~\ref{ssec:provenance}) in the maintenance tier. 

\subsection{Data lake users} 
\label{sec:users}
As shown in Fig.~\ref{fig:rw-high-view}, the complete picture of a data lake also includes 
human users.
Users often interact with a data lake in different roles. 
According to \cite{Chessell2014}, a business data lake scenario typically includes:
(1) \emph{data scientists and business analysts} who build and apply analytics models over the data lake, (2) \emph{information curators} who define new data sources, organize and maintain the metadata in the catalog of existing data sources,  
(3) the \emph{governance, risk, and compliance team} who ensures that the organizational regulations and business policies are followed   (e.g., an auditor), and
(4) the \emph{operations team} that maintains the data lake (e.g., data quality analysts, integration developers). 
Such users can help improve enriching the semantics of data lakes over time, by adding metadata tags and linkage information based on conceptual models or standard vocabularies with respect to ontologies, such as schema.org \cite{pwcdatalake2015, DBLP:conf/birthday/JarkeQ17}.
Moreover, it is not an easy task to design a data lake that has effective control of data security over diverse users and heterogeneous data stores in data lakes. 
CoreDB \cite{DBLP:conf/cikm/BeheshtiBNCXZ17, DBLP:journals/pvldb/BeheshtiBNT18} creates different users or roles for access control, and enables authentication and data encryption.
A few tools are mentioned in \cite{couto2019mapping}  for system authentication, authorization, and data encryption based on the Hadoop platform, e.g., Apache Ranger\footnote{\url{https://ranger.apache.org/}}.  

In this survey, our focus is to conduct a technical comparison of systems, instead of human-system interaction. 
 In the following, we do not emphasize  these different kinds of users and generally call them   data lake users.

\section{Storage}
\label{ssec:DLS}

\begin{table*} [t]
  \centering
  \caption{Classification of  data lake solutions based on functions}
\vspace{-0.2cm}
\label{tbl:all}
\footnotesize 
	\begin{tabular}{|l|l|l|}
		\hline
\textbf{Tier} & \textbf{Functions}            & \textbf{Systems}                                                                                                    \\ \hline   \hline
                                                                               &                                               & GEMMS \cite{quix2016metadata}                                                                                       \\ \cline{3-3} 
                                                                               &                                               & DATAMARAN \cite{DBLP:conf/sigmod/GaoHP18}                                                                           \\ \cline{3-3} 
                                                                               & \multirow{-3}{*}{Metadata extraction}         & Skluma \cite{skluzacek2018skluma}                                                                                   \\ \cline{2-3} 
                                                                               &                                               & GEMMS \cite{quix2016metadata, DBLP:conf/er/HaiQW19}                                                                 \\ \cline{3-3} 
                                                                               &                                               & HANDLE \cite{10.1007/978-3-030-59065-9_7}                                                                          \\ \cline{3-3} 
                                                                               &                                               & Data vault \cite{DBLP:conf/ideas/NogueiraRD18, DBLP:conf/er/Giebler19}                                              \\ \cline{3-3} 
                                                                               &                                               & Diamantiniet al. \cite{DBLP:conf/adbis/DiamantiniGMPSU18, DBLP:conf/sebd/DiamantiniGMPSU18, diamantini2021approach} \\ \cline{3-3} 
                                                                               &                                               & Aurum \cite{DBLP:conf/icde/FernandezAKYMS18}    \\ \cline{3-3} 
\multirow{-9}{*}{Ingestion}                                                    & \multirow{-6}{*}{Metadata modeling}           & Sawadogoet et al. \cite{sawadogo2019metadata}                                                                       \\ \hline  \hline  
                                                                               
                                                                               &                                               & GOODS \cite{DBLP:conf/sigmod/HalevyKNOPRW16, DBLP:journals/debu/HalevyKNOPRW16}                                     \\ \cline{3-3} 
                                                                               &                                               & DS-Prox \cite{DBLP:conf/sisap/AlserafiCA017,DBLP:conf/medi/AlserafiA0C19, DBLP:journals/tois/AlserafiARC20}         \\ \cline{3-3} 
                                                                               &                                               & KAYAK \cite{maccioni2017crossing, maccioni2018kayak}                                                                     \\ \cline{3-3} 
&                                               & Nargesian et al. \cite{nargesian2020organizing}                                                                     \\ \cline{3-3}
                                                                               &                                               & Ronin \cite{ronin10.14778/3476311.3476364}                                                                          \\ \cline{3-3} 
                                                                               & \multirow{-6}{*}{Dataset  organization}      & Juneau \cite{zhang2020finding}                                                                                      \\ \cline{2-3} 
                                                                               &                                               & Aurum \cite{DBLP:conf/icde/FernandezAKYMS18}                                                                        \\ \cline{3-3} 
                                                                               &                                               & Brackenbury et.al.  \cite{DBLP:conf/sigmod/BrackenburyLMEU18}                                                       \\ \cline{3-3} 
                                                                               &                                               & JOSIE \cite{DBLP:conf/sigmod/ZhuDNM19}                                                                              \\ \cline{3-3} 
                                                                               &                                               & $D^3L$ \cite{bogatu2020dataset}                                                                                     \\ \cline{3-3} 
                                                                               &                                               & Juneau \cite{DBLP:journals/pvldb/ZhangI19, DBLP:conf/cidr/IvesZHZ19, zhang2020finding}                              \\ \cline{3-3} 
                                                                               &                                               & PEXESO  \cite{DBLP:conf/icde/DongT0O21}                                                                             \\ \cline{3-3} 
                                                                               &                                               &   RNLIM \cite{10.1007/978-3-030-77385-4_18}                                                                               \\ \cline{3-3} 
                                                                               & \multirow{-8}{*}{Related dataset discovery
}   & DLN  \cite{10.14778/3457390.3457403}
                                                                               \\ \cline{2-3} 
                                                                               &                                               &  Constance \cite{hai2016constance, DBLP:conf/adbis/HaiQZ18, DBLP:conf/er/HaiQK18, hai2019rewrite}
                                                                               \\ \cline{3-3}
                                                                               &  \multirow{-2}{*}{Data integration} & 
                                                                               ALITE \cite{alite}
                                                                               \\ \cline{2-3} 
                                                                               &                                               & CoreDB \cite{DBLP:conf/cikm/BeheshtiBNCXZ17, DBLP:journals/pvldb/BeheshtiBNT18}                                     \\ \cline{3-3} 
                                                                               &                                               & $D^4$ \cite{10.14778/3384345.3384346}                                                                               \\ \cline{3-3} 
                                                                               &                                               & DomainNet \cite{DBLP:conf/edbt/LeventidisRMRG21}                                                                    \\ \cline{3-3} 
                                                                               &                                               & Constance \cite{DBLP:conf/er/HaiQW19}                                                                               \\ \cline{3-3} 
                                                                               & \multirow{-5}{*}{Metadata enrichment}         & GOODS \cite{DBLP:conf/sigmod/HalevyKNOPRW16, DBLP:journals/debu/HalevyKNOPRW16}                                     \\ \cline{2-3} 
                                                                               &                                               & CLAMS \cite{farid2016clams}                                                                                         \\ \cline{3-3} 
                                                                               &                                               & Constance \cite{DBLP:conf/er/HaiQW19}  
                                                                                                                                     \\ \cline{3-3}
                                                                               & \multirow{-3}{*}{Data cleaning}    &  Song et al. \cite{song2021auto}                                                                              \\ \cline{2-3} 
                                                                               & Schema evolution                              & Klettkeet et al. \cite{DBLP:conf/bigdataconf/KlettkeAS0S17}                                                         \\ \cline{2-3} 
                                                                               &                                               & IBM tool \cite{terrizzanodata}                                                                                      \\ \cline{3-3} 
                                                                               &                                               & Suriarachchi et al. \cite{DBLP:conf/eScience/SuriarachchiP16}                                                       \\ \cline{3-3} 
                                                                               &                                               & GOODS \cite{DBLP:conf/sigmod/HalevyKNOPRW16, DBLP:journals/debu/HalevyKNOPRW16}
                                                                               \\ \cline{3-3} 
                                                                               &                                               & CoreDB \cite{DBLP:conf/cikm/BeheshtiBNCXZ17, DBLP:journals/pvldb/BeheshtiBNT18}                                     \\ \cline{3-3} 
\multirow{-29}{*}{Maintenance}                                                 & \multirow{-5}{*}{Data provenance}             & Juneau \cite{DBLP:journals/pvldb/ZhangI19, DBLP:conf/cidr/IvesZHZ19, zhang2020finding}                              \\ \hline  \hline 
                                                                               &                                               & JOSIE \cite{DBLP:conf/sigmod/ZhuDNM19}                                                                        \\ \cline{3-3} 
                                                                               &                                               & $D^3L$\cite{bogatu2020dataset}                                                                               \\ \cline{3-3} 
                                                                               &                                               &  Juneau \cite{DBLP:journals/pvldb/ZhangI19, DBLP:conf/cidr/IvesZHZ19, zhang2020finding}                                                                                      \\ \cline{3-3} 
                                                                               & \multirow{-4}{*}{Query-driven data discovery} & Aurum \cite{DBLP:conf/icde/FernandezAKYMS18}                              \\ \cline{2-3} 
                                                                               &                                               & Constance \cite{hai2016constance, DBLP:conf/adbis/HaiQZ18}                                                          \\ \cline{3-3} 
                                                                               &                                               & CoreDB \cite{DBLP:conf/cikm/BeheshtiBNCXZ17, DBLP:journals/pvldb/BeheshtiBNT18}                                     \\ \cline{3-3} 
                                                                               &                                               & Ontario \cite{DBLP:conf/dexa/EndrisRVA19, handle:20.500.11811/8347}                                                      \\ \cline{3-3} 
\multirow{-8}{*}{Exploration}                                                  & \multirow{-4}{*}{Heterogeneous data querying}    & Squerall \cite{mami2019querying}                                                                                    \\ \hline
\end{tabular}
\vspace{-0.4cm}	
\end{table*}

%
%

An important aspect of a data lake's architecture is the \emph{storage tier}, which specifies the technology used for storing data.
In what follows, we show that some  approaches rely on the common relational or NoSQL databases while others have developed new storage systems, or combinations (polystores); the upper part of Fig.~\ref{fig:rw-high-view} depicts such diverse choices, which could be operated on-premise or in clouds. 
We classify the existing data storage solutions for data lakes by how the ingested data is stored in the  lake: as files 
(Sec.~\ref{ssec:hadoop}), in a single database 
(Sec.~\ref{ssec:singleDB}), or using polystores 
(Sec.~\ref{ssec:poly}). 
We also briefly mention industrial solutions that build data lakes on cloud platforms 
(Sec.~\ref{ssec:dlcloud}).



\subsection{File-based storage systems}
\label{ssec:hadoop}
The Hadoop Distributed File System (HDFS) 
is one of the most frequently mentioned data storage systems for data lakes \cite{Pivotal2013, pwcdatalake2015, BociBigDataArchitecture}. 
HDFS supports a wide range of files \cite{grover2015hadoop}.  
Besides text (e.g., CSV, XML, JSON) and binary files (e.g., images), it   supports certain formats for data compression, e.g., 
Snappy\footnote{\url{https://github.com/google/snappy}}, Gzip\footnote{\url{https://www.gzip.org/}}.
It also allows columnar storage formats such as Parquet\footnote{\url{https://parquet.apache.org/}} and row-based storage format Avro\footnote{\url{https://avro.apache.org/}} that enable easy schema management. 
As one of the most common data storage options for data lake, file systems such as Hadoop are widely used in practice. In this survey, we list a few representative systems as follows. 

Hadoop alone usually does not fulfill the goals of a data lake.  
Microsoft's \emph{Azure data lake store} 
\cite{DBLP:conf/sigmod/RamakrishnanSDK17} offers a hierarchical, multi-tier file-based storage system.\footnote{
	The latest system is known as  \emph{Azure Data Lake Storage Gen2}.}  
It applies \emph{Azure Blob storage}, which is a cloud storage solution optimized for large unstructured object data. 
It also supports HDFS and the Hadoop ecosystem,  
(e.g., Spark, 
Sqoop\footnote{\url{https://sqoop.apache.org/}}). 
Azure also has an indexing subsystem \emph{Hyperspace} \cite{10.14778/3476311.3476382}. 
Another system built upon HDFS is \emph{CLAMS} \cite{farid2016clams}. 
CLAMS is a prototype system that stores the ingested dataset  in HDFS, and allows users to 
register the datasets for constraint discovery and data cleaning for data lakes. 


\subsection{Single data store}
\label{ssec:singleDB}
Some DL systems aim  at specific types of data and employ  
a single database system at their storage tier.
%
As an example, 
the \emph{personal data lake} \cite{DBLP:conf/bdcloud/WalkerA15} applies a  graph-based data model (i.e., property graphs), and stores data in Neo4j. The proposed data lake has a special application focus on user data. Such data is  usually relatively small in size compared to business scenarios, but imposes  higher requirements regarding data privacy. Heterogeneous personal data fragments generated from user-web interaction (structured, semi-structured, unstructured) are serialized to  specifically defined JSON objects. These are flattened to Neo4j graph structures with extensible metadata management in the data lake, categorizing for kinds of data: raw data, metadata, additional semantics, and the data fragment identifiers.

\para{Potential techniques}
Going further, \emph{multi-model databases} support multiple data models and formats in a single database (for a survey, cf.~\cite{lu2019multi}). However, before choosing a multi-model database in a data lake, one should also check the underlying storage strategy, i.e., native storage support for different data models or merely different interfaces to the same storage strategy. If not, polystores should be considered, as discussed next.

\subsection{Polystore systems}
\label{ssec:poly}
\emph{Polystore} (or \emph{multistore}) systems 
provide an integrated access to a hybrid of multiple  data stores for heterogeneous data. 
By definition, a data lake supports heterogeneous data in raw format, e.g., for zone architectures  \cite{2014big}. Thus, polystores are a feasible choice when a data lake is diverse. 

Constance \cite{hai2016constance, DBLP:conf/adbis/HaiQZ18} applies polystore, and stores the diverse raw data according to its original format: relational (e.g., MySQL), document-based (e.g., MongoDB), and graph databases (e.g., Neo4j). 
For instance, a JSON file will be stored in MongoDB.  
If an input dataset cannot be directly stored in a relational or NoSQL store, or considering, e.g., scalability for distributed computing, data can also be stored in HDFS. 
An example would be a data source producing streams of large  binary image files, which requires parallel data compression. 
If these defaults seem inadequate, users can specify the data store via the UI.  
 
 \emph{Google Dataset Search (GOODS)} 
 \cite{DBLP:conf/sigmod/HalevyKNOPRW16, DBLP:journals/debu/HalevyKNOPRW16} supports heterogeneous data storage used in Google's key-value stores \emph{Bigtable} \cite{DBLP:journals/tocs/ChangDGHWBCFG08}, file systems, and \emph{Spanner} \cite{DBLP:journals/tocs/CorbettDEFFFGGHHHKKLLMMNQRRSSTWW13}; see also Sec.~\ref{ssec:dlcloud}. 

Another data lake  service allowing raw data stored in both relational and NoSQL is \emph{CoreDB} \cite{DBLP:conf/cikm/BeheshtiBNCXZ17, DBLP:journals/pvldb/BeheshtiBNT18}. 
To store diverse data from web  applications, besides relational databases (e.g., MySQL, PostgreSQL, Oracle), 
  it supports multiple NoSQL systems, i.e., MongoDB,  HBase\footnote{\url{https://hbase.apache.org/}}, and HIVE. 
JSON is used as a unified format to represent entities.
 
\emph{Juneau} \cite{zhang2020finding} supports data science tasks in data lakes. It mainly focuses on tabular data or nested data that can be easily unnested into relations.
Besides data processed by the notebook kernel, Juneau also handles (Jupyter, JupyterLab, Apache Zeppelin, or RStudio) notebooks, workflows in a notebook, and cells that constitute a workflow.
Moreover, it needs to generate and store the relationships of these data objects as graphs. Thus, it applies both a relational database (PostgreSQL) and a graph database (Neo4j). 

\para{Potential techniques.} Since 2015, the study of polystore has been booming with regard to systems (e.g.,  \emph{Polybase} \cite{polybase},  \emph{BigDAWG} \cite{DBLP:journals/sigmod/DugganESBHKMMMZ15}) 
and open-source tools (e.g., Drill\footnote{\url{https://drill.apache.org/}}, Spark) -- see  \cite{tan2017enabling} for a survey. 
Although they are not claimed as data lakes, they are potentially useful techniques to be  part of a data lake architecture.
In Sec.~\ref{ssec:qrwt}   
   we continue with how to query a polystore. 

 \subsection{Cloud data lakes}
\label{ssec:dlcloud}

%
Recently, it is becoming a  common practice to build large-scale commercial data lakes on cloud infrastructure  \cite{10.1145/3448016.3457546, zagan2021cloud, weintraub2021needle}.  
Cloud-based storage choices include single-cloud, multi-cloud and a hybrid of cloud and on-premise platforms \cite{zagan2021cloud}. 
Several major cloud database vendors are promoting server-less data analytics and native cloud platform for building data lakes, most prominently 
Amazon Web Services (AWS)\footnote{\url{https://aws.amazon.com/big-data/datalakes-and-analytics/}},
Azure Data Lake Store\footnote{\url{ https://azure.microsoft.com/en-us/solutions/data-lake/}},
Google Cloud Platform (GCP)\footnote{\url{https://cloud.google.com/solutions/build-a-data-lake-on-gcp}},  
Alibaba Cloud\footnote{\url{https://www.alibabacloud.com/product/data-lake-analytics}}, and 
 the Data Cloud from Snowflake\footnote{\url{https://www.snowflake.com/workloads/data-lake/}}.
%
%
Cloud platforms have several prominent advantages for  data lakes. 
In a cloud data lake, one can scale storage space and computation power dynamically, and in many cases, the prices of resources are more economical  than on-premises. 
Moreover, major cloud vendors provide many  additional analytics tools in their product portfolio, e.g., AI services and data visualization tools, which make it convenient for developing different
applications 
on top of data lakes.
There are proposals \cite{10.14778/3476311.3476382, weintraub2021needle} on building indexing structures for cloud data lakes. 
Nevertheless, relying on a cloud platform also implies risks and challenges in some aspects such as data  security, data provenance, and fault tolerance. 

\subsection{Summary and discussion}
Choosing the ``right'' data storage system is  one of the most important parts of architecting a data lake. A data lake designer needs to factor in not only the  raw data itself, but also how the data will be used. 
We have shown that the choices are diverse:  file systems or databases (relational or NoSQL), single or hybrid systems, on-premise or cloud, etc. 
The specific choice of storage strategy often shapes the required functions, which we introduce next. 
For future work, it is interesting to see more data lake solutions researched and developed over cloud platforms.

\section{Ingestion}
\label{sec:ing}

During the ingestion phase, a data lake loads raw data. Without any additional information or insights, i.e., \emph{metadata}, a data lake is hardly usable as the structure and semantics of the data are not known; this could potentially turn the data lake into a `data swamp'.
Therefore, it is crucial to acquire as much metadata
as possible from the data sources.
There are different types of metadata: schemata which preserve the structure of the dataset, 
semantic metadata, constraints, and other descriptive information, etc.
During or shortly after data ingestion, existing solutions mainly  extract metadata from the input datasets and model them. 
Thus, in this section, we focus on  metadata extraction (Sec.~\ref{sec:meta_extr})  and modeling  (Sec.~\ref{ssec:mmodel}).

\subsection{Metadata extraction}
\label{sec:meta_extr}
 \emph{Metadata extraction} is the process of discovering 
metadata information of a dataset, often structural metadata is extracted in the first place, but also semantical information and relationships to other datasets is extracted. In a data lake, metadata extraction is essential for accessing datasets in a later phase. 
In Sec.~\ref{sec:metad_enrich},  we further address extracting \emph{hidden} metadata such as functional dependencies.
Given semi-structured or unstructured data,  existing approaches \cite{quix2016metadata, hai2016constance,  DBLP:conf/sigmod/GaoHP18}  extract primarily structural metadata, while in  \cite{skluzacek2018skluma} it extracts metadata related to content and context.

The \emph{Generic and Extensible
	Metadata Management System} (\emph{GEMMS}) for data lakes  \cite{quix2016metadata} is a framework for extracting metadata from heterogeneous sources, which is then stored in an extensible metamodel. 
Since  the data sources and schemata may change over time, 
it is important that the data lake has a flexible and extensible manner of metadata extraction. 
For each input file, GEMMS first detects its format, then initiates a corresponding parser to obtain 
the structural metadata (e.g., trees, tables, and graphs) and metadata properties (e.g., header information implying the content of the file). 
A tree structure inference algorithm is implemented for structural metadata extraction, 
which iterates semi-structured data in a breadth-first manner, and   detects the tree structure. 
A follow-up work, Constance \cite{hai2016constance},  can also extract structural metadata, 
i.e., schemata from semi-structured files such as XML and JSON.

\emph{DATAMARAN} \cite{DBLP:conf/sigmod/GaoHP18} provides a  three-step algorithmic approach to extract structures from semi-structured log files.  The records of its input log files, span multiple lines, with record types and boundaries. It first generates candidate \emph{structure templates}, which use regular expressions \cite{sipser2006introduction} to express the record structure while allowing minor variations;
the structure templates are stored in hash-tables, and only the ones satisfying a coverage threshold assumption are kept. Next, redundant structure templates are pruned based on a specially designed score function, and finally further optimized using two refinement  techniques over the pruned structure templates. The process of DATAMARAN  does not require human  supervision  and provides a high extraction accuracy  compared to existing works. In the experiments, the authors crawled 100 datasets with large log files from GitHub to mimic a real data lake.

\emph{Skluma} \cite{skluzacek2018skluma} extracts   metadata regarding content  and context from  scientific data files. The types of input files are  diverse, e.g., JSON, CSV, unstructured texts, and images. It first finds the name, path, size, and extension  of the files; then it infers file types  and adds specific extractors accordingly to process  tabular data, free texts or null values, etc. With the growing importance of Research Data Management for replicability in scientific studies, 
we expect such  approaches  to grow significantly over the following years.

\subsection{Metadata modeling} 
\label{ssec:mmodel}
\emph{Metadata modeling} answers 
the question of 
how to structure and organize the metadata. 
To make the content of a data lake findable, accessible, interoperable, and reusable (FAIR Principles \cite{FAIR2016}), it is necessary to conduct metadata modeling. The majority of proposed models are either logic-based or graph-structured.
Here we categorize the existing solutions 
based on the \emph{types of  metadata models}: generic models, 
data vault,
and graph based models.

\subsubsection{Generic metadata models}
\label{ssec:gmodel}
The logic-based metadata model of GEMMS \cite{quix2016metadata} has different model elements  and allows the separation of metadata containing information about the content, semantics, and structure. 
It captures the general metadata properties in the form of key-value pairs, as well as structural metadata as trees and matrices to assist querying. 
Moreover, domain-specific ontology terms can be attached to metadata elements as semantic metadata. 
In \cite{DBLP:conf/er/HaiQW19}, this metadata model is extended for representing individual  schemata for relational
tables, JSON, and labeled property graphs of Neo4j.

Another generic metadata model is \emph{HANDLE} \cite{10.1007/978-3-030-59065-9_7}. 
It has three abstract entities:  data, metadata, and property. HANDLE enables flexibility with fine-grained levels, and it adapts the zone architecture mentioned in Sec.~\ref{ssec:pond_zone}. Interestingly, the elements of the GEMMS model can also be mapped to  HANDLE. Finally, HANDLE can be used for linked data and can be implemented in Neo4j.

\subsubsection{Data vault}
For structured or semi-structured data in typical business scenarios, a promising conceptual modeling environment is  \emph{data vault} \cite{DBLP:conf/ideas/NogueiraRD18, DBLP:conf/er/Giebler19}.
It has three main elements types: 
 \emph{hubs} representing business concepts, \emph{links} indicating the many-to-many relationships among hubs, 
 and  \emph{satellites} with descriptive properties of hubs and links \cite{lindstedt2011super, linstedt2015building}.
 %
Nogueira et al. \cite{DBLP:conf/ideas/NogueiraRD18} show how their conceptual model based on data vault   can be transformed into relational and document-oriented logical models, and further to physical models (PostgreSQL and MongoDB, respectively).    
Giebler et al. \cite{DBLP:conf/er/Giebler19} report  
their experience with applying data vaults for data lakes in the domains of manufacturing, finance, and customer service. They also point  out practical obstacles such as inconsistencies among data sources.

\subsubsection{Graph-based metadata models}
\label{sssec:graphmodel}
Adapting ideas about knowledge graphs from the linked data and semantic web communities, several network- or hypergraph-based metamodels have been proposed 
 for data lakes. In the business context,
Diamantiniet al. \cite{DBLP:conf/adbis/DiamantiniGMPSU18, DBLP:conf/sebd/DiamantiniGMPSU18, diamantini2021approach}
propose a network-based metadata model, focusing on business names, data field descriptions, and rules, in addition to data formats and schemata.  It creates a graph-based representation with XML/JSON nodes and labeled arcs indicating their relationship. Nodes can be merged based on lexical and string similarities, and linked to semantic knowledge (e.g., from DBpedia). The authors suggest extracting thematic views of interest to the business, similar to data marts in data warehouses. 
In \cite{diamantini2021approach} the proposed model can support unstructured data, besides (semi-)structured ones.

To efficiently discover relevant datasets from massive data sources, 
\emph{Aurum} \cite{DBLP:conf/icde/FernandezAKYMS18} devises an  \emph{enterprise knowledge graph (EKG)} to capture and query relationships among datasets. An EKG is a hypergraph with three elements: nodes, weighted edges, and hyperedges. 
Nodes represent dataset attributes, which are connected by edges when there is a relationship among them; hyperedges represent different granularities among arbitrary numbers of nodes, e.g., connecting attributes and tables. Aurum builds the EKG,  maintains it upon data changes  and allows users to query it with a graph query language based on discovery primitives. 

Sawadogoet et al.  \cite{sawadogo2019metadata} emphasize six evolution-oriented features of metadata management: semantic enrichment, data indexing, link generation and conservation (discover hidden similarities or integrate existing links among datasets), data polymorphism (preserve multiple transformed forms of the same dataset), data versioning, and usage tracking. 
Taking these features into consideration, their metadata model encompasses the notions of hypergraph, nested graph, and attributed graph. In terms of content, it can describe attributes, objects, datasets, historical versions, (similarity or parent-child) relationships, logs, and indexes.  

\new{
\subsubsection{Summary and discussion}
Metadata extraction and a semantically rich modeling of metadata are crucial issues for the ingestion phase of a data lake. 
The data vault model has been developed for more structured environments like data warehouses and does not seem to fully fit the requirements for unstructured data and flexibility in data lakes. Early lake-specific approaches like GEMMS aimed at structured metamodels (i.e., formally expressed as a UML class diagram), carefully designed to cover all the information possibly relevant for metadata management in a data lake. 
These approaches also have  their limitations in flexibility: the metadata model might be easy extensible, but the management of the metadata 
(i.e.,its storage, user interfaces for creation and manipulation, etc.) is much more challenging. Recent graph-based models are more promising for flexibility, if the  graph-based storage systems are applied for the management of metadata. Furthermore, as the linking of information in organizations and knowledge graphs become more important, graph-based metamodels are a better fit for these techniques.
}


\section{Maintenance}
\label{ssec:maint}
After ingesting heterogeneous raw data from diverse sources, a data lake is a vast collection of unrelated data, for which we have limited information. 
To make the data usable, the data lake needs to further \emph{process and maintain} the raw data, e.g., find more metadata,  discover hidden relationships, and perform data integration, transformation or cleaning if necessary. 
As shown in Fig.~\ref{fig:rw-high-view}, in this section we categorize the maintenance-related functions into seven groups and discuss the corresponding data lake solutions.

\begin{table*}[]
	\centering
	
	\caption{Comparison of DAG-based   dataset   organization approaches in Sec.~\ref{sssec:DAG}} 
	\label{tbl:dag}
	\resizebox{\textwidth}{!}{
\begin{tabular}{|c|c|c|c|c|} 
	\toprule
	\textbf{System}                                                            & \begin{tabular}[c]{@{}c@{}}\textit{KAYAK \cite{maccioni2017crossing, maccioni2018kayak}}\\\textit{(pipeline)}\end{tabular}                       & \begin{tabular}[c]{@{}c@{}}\textit{KAYAK \cite{maccioni2017crossing, maccioni2018kayak}}\\\textit{(task dependency)}\end{tabular}                         & \textit{Nargesian et al.  \cite{nargesian2020organizing}}                                                         & \begin{tabular}[c]{@{}c@{}}\textit{Juneau \cite{zhang2020finding}}\\\textit{(variable dependency)}\end{tabular}                   \\ 
	\hline
	\textbf{Function}                                                          & \begin{tabular}[c]{@{}c@{}}Represent the primitives of a \\data preparation pipeline\end{tabular} & \begin{tabular}[c]{@{}c@{}}Enforce correct execution sequence\\of tasks while parallelization\end{tabular} & Semantic navigation                                                               & \begin{tabular}[c]{@{}c@{}}Measure table relatedness\\w.r.t. notebook workflow\end{tabular}               \\ 
	\hline
	\textbf{Node}                                                             & Primitives                                                                                        & \begin{tabular}[c]{@{}c@{}}Atomic tasks for data \\preparation operations\end{tabular}                     & Sets of attributes                                                                & Notebook variables                                                                                        \\ 
	\hline
	\textbf{Edge}                                                             & \begin{tabular}[c]{@{}c@{}}Sequential execution order\\of two primitives\end{tabular}             & \begin{tabular}[c]{@{}c@{}}Sequential execution order \\of two tasks\end{tabular}                          & Containment relationships                                                         & \begin{tabular}[c]{@{}c@{}}Notebook functions \\(as edge labels)\end{tabular}                             \\ 
	\hline
	\begin{tabular}[c]{@{}c@{}}\textbf{Edge }\\\textbf{direction}\end{tabular} & \begin{tabular}[c]{@{}c@{}}From the previous primitive\\to the subsequent primitive \end{tabular} & \begin{tabular}[c]{@{}c@{}}From the previous task to\\the subsequent task\end{tabular}                     & \begin{tabular}[c]{@{}c@{}}From the superset  to \\the subset  \end{tabular} & \begin{tabular}[c]{@{}c@{}}From the input variable of the \\function to the output variable\end{tabular}  \\
	\bottomrule
\end{tabular}
}
	\vspace{-0.4cm}
\end{table*}

\subsection{Dataset organization} 
\label{ssec:dlorg}
The \emph{dataset organization}  problem studies how to structure and navigate the massive heterogeneous datasets in data lakes. 
Existing solutions in this group of study, 
define new  \emph{structures} to group and organize datasets for better understanding and accessing a data lake. 
We   categorize them based on their \emph{underlying technologies} to construct the structure for a data lake: 
catalogs, 
classification models, 
and directed acyclic graphs. 
Notably, although the following systems all provide  means to organize a data lake, their exact goals differ. For example, KAYAK  \cite{maccioni2017crossing, maccioni2018kayak} and   Juneau  \cite{zhang2020finding} serve for the purpose of data science, while DS-Prox \cite{DBLP:conf/sisap/AlserafiCA017,DBLP:conf/medi/AlserafiA0C19, DBLP:journals/tois/AlserafiARC20} is a pre-filtering step of schema matching \cite{rahm2001survey}.

\subsubsection{Catalog-based organization} 
\label{sssec:catalog}
 GOODS   \cite{DBLP:conf/sigmod/HalevyKNOPRW16, DBLP:journals/debu/HalevyKNOPRW16} allows datasets to be created, stored, and modified first, before conducting metadata collection. 
For each dataset, it collects various metadata and adds it as one entry in the \emph{GOODS catalog}, which  is stored in Bigtable.
To organize, profile  and search datasets
(e.g., cluster different versions of the same dataset), 
the metadata is classified into six categories, including basic, content-based, provenance, user-supplied, team, project, and temporal metadata. 
This categorization of metadata is closely related to Google's specific information retrieval requirements. 
If a 
data lake developer applies similar catalog-based organization strategy, she
should be encouraged to customize the metadata catalog  to her own needs.

\subsubsection{Classification model based organization}
\label{sssec:dsprox}
 \emph{DS-Prox} \cite{DBLP:conf/sisap/AlserafiCA017} and a later version \emph{DS-kNN} \cite{DBLP:conf/medi/AlserafiA0C19} 
consider the dataset organization problem as a classification problem. 
In specific, DS-kNN incrementally adds every dataset into a new or existing category by applying k-nearest-neighbour (k-NN) search.  
Before the step of classification, DS-kNN first conducts data preparation by feature extraction.
For each attribute, depending on whether its values are  continuous or discrete, DS-kNN
extracts statistical or distribution-based  features respectively, e.g., average numeric mean, or the average number of values. Such data-based features are added to each dataset, together with other features based on extracted metadata, e.g., the number of attributes, and types of each attribute.
Using these features, DS-kNN computes dataset similarity by employing Levenshtein distance \cite{manning2009probabilistic}. 
Next, given a new dataset, 
the proposed classification-based algorithm returns top-k neighbors (classified datasets), 
from which DS-kNN chooses the most frequently appeared category, then assigns the current dataset to this category; 
if none of the existing datasets are found, the new dataset is assigned to a new category. 
Finally, the datasets in the lake can be visualized as a graph: each node is a dataset, and edges between two nodes are labeled with the similarity of the two datasets. 
A later work \cite{DBLP:journals/tois/AlserafiARC20} uses supervised ensemble models   to obtain the similarity values between  dataset pairs. 



\subsubsection{DAG-based organization}
\label{sssec:DAG}
In what follows, we introduce the dataset organization solutions that apply \emph{directed acyclic graphs (DAGs)}. Although they all apply DAGs to organize and navigate a data lake, the exact functions and the  definitions of DAGs differ, as listed in Table~\ref{tbl:dag}.

\emph{KAYAK} \cite{maccioni2017crossing, maccioni2018kayak} aims to support data science pipelines in data lakes, and it organizes the dataset relationships and operations. 
It defines a data lake as a collection of datasets, 
and manages the operations on the datasets, which are the basic blocks of data preparation pipelines.  
\emph{Data preparation} refers to the processing of raw data, so that it can be used in downstream tasks, e.g., analytics. 
KAYAK first defines atomic tasks such as basic profiling and dataset joinability computation.  
Then a sequence of such  atomic tasks further builds up a specific operation for data preparation, referred to as a \emph{primitive}, e.g., insert a dataset. 
 Table~\ref{tbl:dag} shows two different usages and definitions of DAGs in KAYAK, pipeline and task dependency.
To represent data preparation pipelines,    it uses a   DAG  with primitives as nodes and their dependencies (based on execution order) as edges.  
%
To  manage  dependencies among tasks and execute the atomic tasks of a primitive in parallel, 
 KAYAK defines   the second type of DAG for task dependency as shown in Table~\ref{tbl:dag}. 
Here each node represents an atomic task,  
and the directed edges indicate the execution order of two tasks. 
Such a DAG helps to identify which tasks can be parallelized  during execution.
 
%

Nargesian et al. \cite{nargesian2020organizing} define the \emph{data lake organization} problem as discovering the optimal structure to effectively find the desired dataset in a data lake. Such a structure for navigating data lakes is referred to as an \emph{organization}: 
\begin{itemize}[leftmargin=*]
	\item 
As listed in Table~\ref{tbl:dag}, a DAG-based   organization in \cite{nargesian2020organizing}, has sets of attributes as nodes.
The leaf nodes are attributes of input tables, while
non-leaf nodes have  a topic label that summarizes the set of attributes or topics represented by its child nodes. 
The edges represent containment relationships between the set of attributes represented by the nodes. 
\item  To measure semantic similarities among attributes, attribute values are associated with n-dimensional representations \cite{Nargesian2018}, which enable the use of cosine similarity. 
The process of navigation is formalized as a \emph{Markov model}, where the states are the nodes (i.e., sets of attributes) and transitions are the edges, i.e., future states depend only on the current state, not on all the historical states. 
Thus, given a query asking about a topic (e.g., searching keyword is $food$), the transition probability depends only on the current node in the DAG and the similarities between its child nodes and the given topic. The proposed algorithms in   \cite{nargesian2020organizing} try to find the organization structure that achieves the maximum  probability for all the attributes of tables to be found. 
\end{itemize}
A more recent system 
\emph{RONIN} \cite{ronin10.14778/3476311.3476364}, combines navigation using the above DAG-based  structure \cite{nargesian2020organizing}, with   metadata keyword search and joinable dataset search in a data lake.

In Sec.~\ref{ssec:poly} we have mentioned that 
\emph{Juneau} \cite{zhang2020finding}  handles  computational notebooks, workflows, and cells, from which  it builds graphs for data management.
A \emph{workflow graph} is a directed bipartite graph with two types of nodes: \emph{data object nodes} which represent input/output files or formatted text cells, and \emph{computational module nodes} representing code cells in a Jupyter notebook. 
If the data object is the input or output of the computational module, there is a directed edge connecting their nodes.
%
Moreover, as shown in Table~\ref{tbl:dag}, Juneau also has a DAG  for  managing the relationships  of variables in notebooks, referred to as \emph{variable dependency graphs}. 
In a variable dependency graph, nodes represent the variables, and the labeled, directed edges indicate that one variable is computed using another variable through a function. 
Via subgraph isomorphism, Juneau is able to discover tables sharing similar workflows of notebooks (similar sequences/patterns of variables and functions).

\new{\subsubsection{Summary and discussion}
In this subsection, we have discussed various methods on organizing datasets.
Yet, each of these methods comes with its own use cases, merits and limitations.}

\new{In particular, while GOODS \cite{DBLP:conf/sigmod/HalevyKNOPRW16, DBLP:journals/debu/HalevyKNOPRW16} presents an innovative way to build a dataset metadata catalog, it is heavily tailored to datasets and practices used to produce them in Google; the majority of metadata crawled are connected to standardized processes used inside the company to create, maintain and store datasets. }
\new{Methods based on classification models, like DS-Prox \cite{DBLP:conf/sisap/AlserafiCA017} and DS-kNN \cite{DBLP:conf/medi/AlserafiA0C19}, are ideal when the goal is to group datasets sharing some kind of relatedness. 
 Yet, there are types of relatedness among datasets that might not be covered by such simple metadata  based on data instances, e.g., semantics-aware dataset unionability \cite{Nargesian2018}.} 

\new{Finally, DAG-based organization methods vary considerably in terms of functionality. 
KAYAK \cite{maccioni2017crossing, maccioni2018kayak} computes inter-dataset metadata regarding only equi-joins among tabular datasets. 
Moreover, Nargesian et al.  \cite{nargesian2020organizing} and RONIN \cite{ronin10.14778/3476311.3476364} focus on organizing attributes as DAG to maximize the probability of users finding relevant tables with respect to their needs. Importantly, this organization is based on containment similarities among the attributes, which means that fuzzy-kind similarities are not supported. 
On the other hand, Juneau \cite{zhang2020finding} exploits workbook metadata, which presents a promising signal of inter-dataset similarity.} 
\subsection{Related dataset discovery}

\label{sssec:relDS}

In data lakes, the process of  \emph{related dataset discovery}, also referred to as \emph{data discovery},   tries to find a subset of relevant datasets  that are similar or complementary to a given dataset in a certain way, e.g., with similar attribute names or overlapped instance values. 
Since a data lake stores and manages a large number of datasets, 
it is neither realistic nor necessary to query or integrate all of them. Therefore, it can be more beneficial to first discover datasets that are useful for a specific purpose.  
Moreover, the relatedness discovered among datasets is also a  valuable and essential type of metadata for exploring a data lake and preventing a data swamp, e.g., for enabling entity resolution or resolving inconsistency across datasets.



As shown in Table~\ref{tbl:dd}, we  present  the solutions
that address the related dataset discovery problem in data lakes.
 The systems in this group mainly handle tabular data, or hierarchical data that can be transformed into tabular data (not necessarily relational data, i.e., some may even violate the first normal form). 
 We categorize these systems primarily based on the \emph{types of relatedness} they use:   joinable tables \cite{DBLP:conf/icde/FernandezAKYMS18, DBLP:conf/sigmod/BrackenburyLMEU18, DBLP:conf/sigmod/ZhuDNM19, bogatu2020dataset} (Sec.~\ref{ssec:early}), tables related for data science tasks \cite{DBLP:journals/pvldb/ZhangI19, DBLP:conf/cidr/IvesZHZ19, zhang2020finding} (Sec.~\ref{ssec:dds}), and tables with semantic relationships \cite{DBLP:conf/icde/DongT0O21, 10.1007/978-3-030-77385-4_18} (Sec.~\ref{ssec:sdsr}). 
 The fourth group of approaches focuses on the \emph{scalability} issue during the discovery process \cite{10.14778/3457390.3457403} (Sec.~\ref{sssec:dln}).
 However, note  that solutions belonging to one category might also be  applicable to other cases: joinable tables can be used for data science, and semantically related tables could also be joinable, etc. 

\begin{table*}[tb]
	\centering
		\caption{Comparison of  related dataset discovery approaches in data lakes}
	
	\label{tbl:dd}
	 \vspace{-0.2cm}
\begin{tabular}{|l|l|l|l|} 
	\hline
	\textbf{Systems}             & \textbf{Relatedness criteria}                                                                                                                                                                                                          & \textbf{Similarity metrics}                                                                                        & \textbf{
Applied technique}                                                       \\ 
	\hline
	\textit{Aurum} \cite{DBLP:conf/icde/FernandezAKYMS18}              & \begin{tabular}[| X | c |]{@{}l@{}}Instance value overlap~\\Attribute name~~\\PK-FK candidate~~\end{tabular}                                                                                    & \begin{tabular}[c]{@{}l@{}}Jaccard similarity \\(MinHash)\\Cosine similarity \\(TF-IDF)\end{tabular} & Hypergraph \\ 
		\hline
	\textit{Brackenbury et.al.}  \cite{DBLP:conf/sigmod/BrackenburyLMEU18} & \begin{tabular}[c]{@{}l@{}}Instance value overlap~~\\Attribute name \\Semantics\\Descriptive metadata~\end{tabular}                                                                                                            & \begin{tabular}[c]{@{}l@{}}Jaccard similarity \\(MinHash)\end{tabular}                                             & -                                                                                  \\
		\hline
	\textit{JOSIE} \cite{DBLP:conf/sigmod/ZhuDNM19}              & Instance
	value overlap~                                                                                                                                                                                                        & Intersection size of sets                                                                                          & Inverted Index 
 \\ 
	
	\hline
	$D^3L$  \cite{bogatu2020dataset}                     & \begin{tabular}[c]{@{}l@{}}Instance~ value~ overlap \\Attribute~ name~\\Semantics\\Data value representation pattern~\\(Numerical) data distribution\end{tabular}                                                               & \begin{tabular}[c]{@{}l@{}}Jaccard similarity \\(MinHash)\\\\Cosine similarity \\(Random projections)\end{tabular} &
	5–dim Euclidean space                                                               \\ 
	\hline
\textit{Juneau} \cite{DBLP:journals/pvldb/ZhangI19, DBLP:conf/cidr/IvesZHZ19, zhang2020finding}
& \begin{tabular}[c]{@{}l@{}}Instance value overlap\\Domain overlap\\Attribute name\\Key constraint\\New attributes rate\\New instance rate\\Variable dependency\\Descriptive metadata\\Null Values\end{tabular}                                    & Jaccard similarity                                                                                                 & \begin{tabular}[c]{@{}l@{}}Workflow graph\\Variable dependency graph\end{tabular}                                                    \\ 
\hline
\textit{PEXESO}  \cite{DBLP:conf/icde/DongT0O21}                                                                                                                                   & (Textual) instance values                                                                                                                                                                                                                                                                                                                                                                                                  & \begin{tabular}[c]{@{}l@{}}Any similarity function\\in a metric space\end{tabular}                                & \begin{tabular}[c]{@{}l@{}}High-dimensional vectors \\Hierarchical grids\\Inverted Index\end{tabular}  \\ 
\hline
\textit{RNLIM} \cite{10.1007/978-3-030-77385-4_18} & \begin{tabular}[c]{@{}l@{}}Table name\\Attribute name\\Attribute data type\\Attribute value domain\end{tabular}          & --                                                                                                                 & BERT \cite{devlin2019bert}                                           \\
\hline
\textit{DLN} \cite{10.14778/3457390.3457403}                                                                                                                                         & \begin{tabular}[c]{@{}l@{}}Attribute name\\Instance values\end{tabular}                                                                                                                                                                                                                                                                                                                            & \begin{tabular}[c]{@{}l@{}}Jaccard similarity\\Cosine similarity \end{tabular}                                 & Classification models                                                                                                                 \\ 
\hline

\end{tabular}

  \vspace{-0.4cm}	
\end{table*}

\subsubsection{Discovery of joinable datasets}
\label{ssec:early}

\emph{Aurum} \cite{DBLP:conf/icde/FernandezAKYMS18}  enables the discovery of joinable datasets by building a
hypergraph (i.e., EKG) which stores information on how columns of different tabular datasets might be related.
To construct it
, Aurum first profiles each table column by adding \emph{signatures}, i.e.,  information extracted from column values such as cardinality, data distribution, and  a representation of data values (i.e., MinHash). 
Then, it indexes these signatures using locality-sensitive hashing (LSH). 
When two columns have their signatures indexed into the same bucket after hashing, an edge is created between corresponding nodes, and their similarity score is stored as the edge weight. Aurum also detects primary-foreign key relationships between columns by first inferring approximate key attributes. 
A highlight of Aurum is its efficiency of computing set similarities. More specifically, given the total number  $n$ of attributes of all datasets, instead of conducting an all-pair comparison of $\mathcal{O}({n}^2)$ complexity, it profiles  columns with signatures and stores them in an LSH-index; then, by using approximate nearest neighbor search, it reduces to linear complexity. 
When  changes occur in the data,  Aurum does not re-read it from scratch. Only if the difference compared to the original values is above a threshold, it updates  column signatures and the hypergraph.

Brackenbury et al.  \cite{DBLP:conf/sigmod/BrackenburyLMEU18} provide a  high-level data lake proposal, 
which  shares a similar idea to Aurum, in terms of using multiple criteria to measure dataset similarities. The difference is that when the algorithms alone cannot provide reliable suggestions, it also includes humans in the loop. 
To find joinable datasets, it measures the similarity of files (e.g., HTML tables), and considers approximate matches in terms of data values, schemata and descriptive metadata (the source of data, information added by users, etc.). For measuring the similarity of the files and clustering them, it 
computes the Jaccard similarity between file paths using  MinHash  and  LSH.

 
 \emph{JOSIE} \cite{DBLP:conf/sigmod/ZhuDNM19} handles data lakes with tabular data, e.g., a corpus of web tables.
 It addresses two challenges with regard to applying existing overlap set similarity search solutions in data lakes: \emph{i}) a data lake may contain a large number of tables; hence the number of columns and distinct values could also be large, and \emph{ii}) it could be difficult for a human user to directly give an appropriate threshold value $\theta $ for the intersection value. 
Thus,  an exact top-k overlap set similarity search approach is proposed in \cite{DBLP:conf/sigmod/ZhuDNM19}, which enables \emph{i}) scaling to large sets (with size over 1K and maximum size in the
millions) and \emph{ii}) returning top-k results without the need of human-defined threshold value $\theta $.

%
Given a table  $T$ in the data lake, and one specific column $C$ from $T$, JOSIE can return tables in the data lake that could be joined with $T$ on $C$. The task is formalized as the problem of \emph{overlap set similarity}, which considers the table columns as sets, and the same tuple values as the set intersection. 
Each table in the output, contains a column that has an overlap with  $C$, and the intersection value is larger than a given threshold $\theta $. 
Then naturally, the problem of joinable table discovery is transformed into the problem of finding the exact top-k overlap set similarity search. 
The measurement used in JOSIE is the  \emph{intersection size} of the sets, also referred to as \emph{overlap similarity}. 
For returning top-k sets JOSIE has applied  \emph{inverted indexes}, which map between the sets and their distinct values and make  
JOSIE   scalable with a large number of tables. 
JOSIE employs a cost model to eliminate the unqualified candidates effectively. Such a method makes the performance robust to different data distributions.

$D^3L$ \cite{bogatu2020dataset} also incorporates multiple criteria to decide whether a dataset is relevant to another. In particular, it regards five signals of dataset similarity: \emph{i}) attribute name similarity, \emph{ii}) instance value overlaps between columns, \emph{iii}) embedding similarity of columns, \emph{iv}) format similarity of instance values, and \emph{v}) distribution similarity of numerical attributes. 
%
Therefore, given table  attributes,  $D^3L$ first transforms schemata  and data instances to intermediate representations of  q-grams, TF/IDF tokens, regular expressions, word-embeddings \cite{joulin2017bag}, and the Kolmogorov-Smirnov
statistic \cite{conover1998practical}. 
Based on these five features, $D^3L$ transforms the problem of finding the relatedness between tables to the calculation of weighted Euclidean distance in a 5–dimensional space. In doing so, the weight of each feature (i.e., feature coefficients) indicates its significance for the combined distance.
To tune the feature weights, $D^3L$ trains a \textbf{binary classifier} over a training dataset with relatedness ground truth, and applies the coefficients of the trained model as the weight of features for distance calculation.
Similar to Aurum, $D^3L$ builds LSH to index the features and maps them to the distance space, where two items are considered to be similar if they are hashed into the same bucket. An interesting finding is that using LSH to discover joining paths leads to accurate discovery of more related tables (and attributes).

 \subsubsection{Discovery of task-specific datasets for data science }
\label{ssec:dds}

Juneau \cite{DBLP:journals/pvldb/ZhangI19, DBLP:conf/cidr/IvesZHZ19, zhang2020finding} provides searching over related tables from a different perspective. 
It extends computational notebooks (e.g., Jupyter) and supports common data science tasks, such as finding additional data for training or validation, and feature engineering.
When users specify the desired target table,
the system can automatically return a ranked list of tables, 
which might  be relevant to the given table. 
Specifically, as shown in Table~\ref{tbl:dd}, Juneau extends the notion of ``relatedness'' with the following signals. 
 \begin{enumerate}[leftmargin=*]
 	\item  

In addition to the instance value overlap  and similar attribute names,  it considers pairwise matched attributes that share similar domains, and matched candidate key pairs.  
The proposed similarity metrics are based on Jaccard similarity; sketches and LSH-based approximation \cite{DBLP:conf/icde/FernandezMNM19,zhu2016lsh} are mentioned as alternatives for scalability.

 	\item To augment user queries, it may suggest data instances or attributes in the candidate tables, which do not exist in the target table.

 	\item 
Based on the variable dependency graphs (Sec.~\ref{ssec:dlorg}), it defines the \emph{provenance similarity} of two tables based on the graph similarity of their variables. This measurement aims to help connect variables and tables via user-defined workflow operations. This allows   finding new tables that are related to the current table via workflows. 

	\item 
 Juneau also identifies similar tables with regard to descriptive metadata 
(e.g., information about the data science task),
  and the number of null values (e.g., fill missing values in a data cleaning task). 
 \end{enumerate}

For a specific data science task, Juneau picks a subset of  relatedness features and computes similarities based on them. For instance, when searching tables for a data cleaning task, it considers the instance value overlap, schema overlap, provenance similarity, and null value differences.

\subsubsection{Discovery of semantically related datasets}
\label{ssec:sdsr}

PEXESO \cite{DBLP:conf/icde/DongT0O21} tackles the problem of finding semantically joinable tables when considering only textual attributes. Towards this direction, it transforms  textual  values into high-dimensional vectors, and computes their vector similarities. 
For efficient similarity computation among such representation vectors, it utilizes an inverted index, and a \emph{hierarchical grid} which is used for partitioning the space.

The \emph{Relational Natural Language Inference Model} (RNLIM) \cite{10.1007/978-3-030-77385-4_18} is a framework that transforms related attribute discovery to unsupervised natural
language inference \cite{maccartney2008phrase}, which determines whether a  hypothesis can be inferred from the given premise texts.   
Contrary to other data discovery systems, it focuses on specifying semantic relationships between the tables. That is, given a pair of attributes, RNLIM optimizes a neural network for labeling their relatedness. 
More specifically, 
RNLIM considers four signals and separates them into two groups: table and attribute names, attribute data types and attribute value domains. For each such group, it uses multiple matching methods. 
For instance, to perform the domain match between numerical attributes, it uses the Kolmogorov–Smirnov  statistic, which is similar to $D^3L$ \cite{bogatu2020dataset}. 
Using pre-trained language representation models from BERT \cite{devlin2019bert}, RNLIM generates similarity–preserving representations from these two groups of signals, which enable the training of a classification model.

\subsubsection{Scalable related dataset discovery}
\label{sssec:dln}
Data Lake Navigator (DLN)  \cite{10.14778/3457390.3457403} has a different focus compared to the aforementioned data discovery systems, which often require   processing of all the available data, and hence hinder scalability.
DLN tackles the problem of handling large-volume data at the enterprise level, e.g., a data lake with petabytes or even exabytes of data. 
Consider a  data lake with stream data. 
DLN discovers related columns in the streams with respect to a given column.  
The core solution of DLN is building random-forest classification models. 
In specific, DLN considers textual and numerical attributes, and extracts two types of features from them: metadata features, including attribute names and uniqueness,  
and data-based features.
Accordingly, it builds two classifiers. The first classifier uses only metadata features. The second classifier is an ensemble model, which only uses metadata features for numeric attributes, and  both metadata features and data features for textual attributes. 
Notably, for learning classification models DLN needs labeled samples. In essence, it labels the attribute-pairs in the JOIN clauses of queries as positive samples (related columns), whereas it samples negative examples of attribute pairs that never appear in any JOIN clause (non-related columns).

\subsubsection{Summary and discussion} 

Related dataset discovery is a well-researched topic with respect to data lakes. Yet, among all the different methods that have been proposed, one can identify a standard procedure that they follow: 
the first step is to define and extract relatedness signals from tables w.r.t. data (e.g., value overlaps, data distribution patterns), schemata (e.g., attribute names, key constraints), semantics, and descriptive metadata. 
The next step is to compute multi-dimensional similarities between attributes (e.g., based on  Jaccard similarity or
cosine similarity), and aggregate them to an overall similarity between tabular datasets. 
The LSH index and its extensions (e.g., LSH Forest \cite{bawa2005lsh}) are often used to index and map feature values to boost performance or increase the accuracy of relatedness. 
Another important part of data discovery is querying the data lake, which is discussed in Sec.~\ref{ssec:qds}.

%
Nonetheless, we also find that data discovery solutions may differ in their focus.
 Aurum is fast and robust against data value changes and offers a graph-based structure. 
 JOSIE shows a high performance. $D^3$L improves the accuracy of discovered related tables by  dimensions of similarities. 
 Juneau emphasizes workflows for multiple data science tasks. 
 To obtain the semantic relatedness, PEXESO uses high-dimension vectors, while RNLIM relies on BERT.
DLN addresses the challenge of related dataset discovery for  exabyte-scale data lakes.
 Therefore, as shown in Table~\ref{tbl:dd}, their individual relatedness criteria, similarity measures, and applied techniques vary significantly. 
Recent system demonstration proposals also indicate the possibilities of applying knowledge graphs \cite{10.14778/3476311.3476317} and example-based interaction \cite{10.14778/3476311.3476353} for data discovery in data lakes.

Data discovery has been intensively studied beyond the scope of data lakes. We refer the reader to recent tutorials \cite{nargesian12data} and surveys \cite{zhang2020web} for exploring more \textbf{potential} solutions for data discovery in data lakes.
It is our firm belief that data discovery solutions for data lakes will continue being introduced, due to the value and insights they bring to businesses and organizations.

\subsection{Data integration}
\label{ssec:di}
\emph{Data integration (DI)} studies the problem of combining multiple heterogeneous   data sources 
and providing   unified data access for users \cite{doan2012principles}. 
Given a large scale of sources in a data lake, 
users might need to first discover a subset of relevant datasets, before 
resolving the heterogeneities of sources  with regard to data models and schemata. Thus, in some literature \cite{DBLP:journals/pvldb/Miller18}, the related dataset discovery (cf. Sec.~\ref{sssec:relDS}) is also considered as part of data integration. Here we consider the fundamental data integration steps include schema matching \cite{rahm2001survey}, schema mapping \cite{fagin2009clio},  entity linkage \cite{brizan2006survey}, query reformulation \cite{10.1007/s007780100054}, etc. 


Few data lake proposals provide an end-to-end data integration pipeline.
Constance \cite{hai2016constance} uses the generic metadata model for extracted schemata of relational, JSON documents and graph data (see Sec.~\ref{ssec:gmodel}).  
For data integration Constance  first performs schema matching, which finds semantically related attributes.
Users can  select a subset of data sources and schema elements via the user interface, and the system generates an \emph{integrated schema}  for   \emph{partial} integration.
Next, Constance generates schema mappings, which preserve the relationships between the source schemata and integrated schema \cite{DBLP:conf/er/HaiQK18}.  
With schema mappings Constance performs query rewriting and data transformation in a \textbf{polystore}-based setting \cite{DBLP:conf/adbis/HaiQZ18}.
It rewrites the input user query (against the integrated schema) to  subqueries (against source schemata), executes the generated subqueries in the query languages of each data store (e.g., MySQL, MongoDB, Neo4j), and retrieves the subquery results. 
For the final integrated results it further resolves the data type and value conflicts while merging the subquery results.
It also pushes down selection predicates to the data sources to optimize query execution and  reduce the amount of data 
to be loaded. 

\new{\emph{ALITE} \cite{alite} deals with the problem of integrating related tables in data lakes that have been obtained from dataset discovery tasks (Sec. \ref{sssec:relDS}). Particularly, the method gathers results from top-k unionable and joinable queries on datasets and applies holistic schema matching. 
To do so, it leverages embeddings on language models, namely distributed vector representations with the following property: data that are similar to each other are embedded close in the dimensional space (based on the distributional hypothesis \cite{harris1954distributional}). Specifically, ALITE embeds columns by using state-of-the-art techniques such as TURL \cite{deng2022turl}, and then applies hierarchical clustering in order to obtain sets of columns that are related. Finally, based on the aligned columns, it computes the Full Disjunction (FD) \cite{galindo1994outerjoins} among discovered datasets in an optimized way.}


\subsection{Metadata enrichment}
\label{sec:metad_enrich}
In this survey, we refer to \emph{metadata enrichment} as the process of creating implicit metadata from raw data in the data lake, which often requires intensive computation or human effort. 
Notably, in  Section~\ref{sec:meta_extr} we have discussed   systems \emph{extracting} embedded metadata. 
\new{
Here we discuss the necessity to compute and extract ``more hidden'' information from the data, which helps to better understand and explore  datasets in the data lake. 
Such a process is time-costly, and sometimes impossible or unnecessary to be conducted during data ingestion. The metadata discussed in this section is more relevant to the functions in the maintenance tier. 
For example, the semantic metadata enriched by CoreDB can be used for data provenance. The semantic information of domains extracted by $D^4$ and DomainNet could be used to improve the process of related dataset discovery. Structural metadata discovered by Constance can be used for data cleaning, while descriptive metadata enriched by GOODS can be used for dataset organization and data provenance.
}

Next, we discuss systems 
fulfilling such a goal here, categorizing them based on the \emph{types of metadata} that they discover: 
semantic, 
or descriptive metadata. 

\subsubsection{Semantic metadata enrichment} CoreDB \cite{DBLP:conf/cikm/BeheshtiBNCXZ17, DBLP:journals/pvldb/BeheshtiBNT18} is a data lake service that aims at extracting insights from raw data. 
It first extracts essential information representative of the original raw data, referred to as features, e.g., keywords and named
entities. 
Then it provides services that add  synonyms and stems to such features, while it
connects them to open knowledge bases such as Google Knowledge Graph\footnote{\url{https://developers.google.com/knowledge-graph/}}, Wikidata\footnote{\url{https://www.wikidata.org/wiki/Wikidata:Main_Page}}. 
CoreDB also annotates and groups the   data sources in the data lake. 
 
$D^4$ \cite{10.14778/3384345.3384346} tackles the problem of \emph{semantic type detection}, also known as \emph{domain discovery}. That is, 
given a set of input tables, $D^4$ discovers their semantic domains and represents each domain with a set of terms. 
For instance, if there are several color-related attributes, e.g., $vehicle\_color$, $building\_color$, $cloth\_color$, then one of the output domains of $D^4$ is $color$, and it is represented by terms $\{ red, white, black, green, \dots \}$. 
The complete list of the terms of a domain, may come from multiple attributes, while an attribute may contain terms for several different domains. 
$D^4$ applies a data-driven approach, i.e., it processes all the data in the given set of datasets. 
Additionally, the approaches applied in $D^4$ allow it to cope with a large number of tables and attributes, and ambiguous terms (e.g., \emph{Apple} can be a type of fruit or a brand name).  

\emph{DomainNet} \cite{DBLP:conf/edbt/LeventidisRMRG21} tackles a similar problem as $D^4$. It also discovers hidden semantics, and handles the ambiguity and incompleteness in data values. 
For instance, when the value \emph{Apple} appears in multiple tables of a data lake, DomainNet tries to find out if it represents the semantics of one domain (fruit or brand), or both. 
As an approach developed for data lakes, it assumes that a priori knowledge about datasets could be missing, like the types of entities in a table.
Its proposed approach includes building a network graph using data values and attribute names, followed by applying community detection over such a network.

\subsubsection{Structural metadata enrichment} 
\label{ssec:rfd}
Constance \cite{DBLP:conf/er/HaiQW19}  enriches the metadata in the data lake by   discovering  \emph{relaxed functional dependencies (RFD)} \cite{caruccio2016relaxed}. 
The relaxed functional dependencies are relaxed in the sense that they do not apply to all tuples of
a relation, or that  similar  attribute values are also considered to be matched.
Such dependencies provide insights that specific  attributes functionally depend
on some other attributes in a loose manner, which apply to the ingested datasets even though they have a certain percentage of inconsistent tuples. 

\subsubsection{Descriptive metadata enrichment} In order to obtain metadata that describes dataset origin, ownership, and its possible usage, 
it is often beneficial to keep human experts in the loop. 
Google's data lake 
GOODS \cite{DBLP:conf/sigmod/HalevyKNOPRW16, DBLP:journals/debu/HalevyKNOPRW16} stores metadata of its datasets in the catalog, 
and it applies \emph{crowdsourcing} for metadata enrichment. 
For instance, it allows adding descriptive metadata of datasets, marking datasets worth additional security attention, such that
people from different teams of the organization (e.g., data owners, auditors, users) can  exchange and communicate about the information of the datasets. 


\subsection{Data cleaning}
\label{ssec:dq_improv}
\emph{Data cleaning} is the process of discovering and fixing data quality problems. 
The data quality problems may reside in one or multiple sources, at the schema level or the instance level \cite{rahm2000data}. For example, a dataset may have missing values, misspellings and redundancies in its instances.
  When we talk about data cleaning in data lakes, we refer to dealing with data quality issues residing in the ingested raw data. 
Given the volume and variety of the data in a lake, it is ideal that the data cleaning approach can work with  heterogeneous raw data, and reduce human effort. 
Thus, there are certain proposals about how to obtain hidden ``rules'' from the data in the data lakes, and then use them to  improve the data quality. 
We divide the systems  in this group based on the methods applied: constraint inference 
or validation rule inference. 

\subsubsection{Data cleaning by constraint inference} 
CLAMS \cite{farid2016clams}  uses \emph{conditional denial constraints} to detect the potentially erroneous data. 
Given the RDF triples, a conditional denial constraint specifies a set of negation conditions about the tuples. 
The proposed approach automatically detects such constraints by discovering possible schemata from RDF data, and corresponding constraints. 
It examines the triples violating the obtained constraints 
and uses them to build a hypergraph, which indicates the number of constraints  violated by each triple. 
Then, it accordingly ranks the RDF triples and asks the user to validate whether such a candidate dirty triple should be removed. 

Constance \cite{DBLP:conf/er/HaiQW19} also uses discovered dependencies for data cleaning, whereas it applies  \emph{relaxed  functional dependencies}.
These dependencies are especially useful 
in cases where the source data has lower quality with inconsistencies and incorrect values.
By using relaxed  functional dependencies, Constance identifies the data objects violating the detected  dependencies, which could be potentially erroneous data. 

\subsubsection{Data cleaning by  validation rule inference}
In \cite{song2021auto}, Song et al. have tackled a specific data cleaning problem, i.e., data validation. In a large enterprise data lake with terabytes of data, the data may change with time. 
The data validation rules indicate whether the changes are significant enough, and will affect the downstream applications. 
The approach in \cite{song2021auto} tries to automatically derive such rules from the machine-generated, string-valued data,  rather than inferred by human experts. 
In principle, it formulates the rule inference problem as an optimization problem, which balances between   false-positive-rate minimization and quality issue preserving.

\subsection{Schema evolution}
\emph{Schema evolution} requires handling the changes of schemata and integrity
constraints \cite{curino2013automating}.
In data lakes, the possible challenge of schema
evolution could be the heterogeneity of the schemata
and the frequency of the changes. While data
warehouses have a relational schema that is
usually not updated very often, data lakes are more
agile systems in which data and metadata can
be updated very frequently. 

Klettke et al. \cite{DBLP:conf/bigdataconf/KlettkeAS0S17} address  the problem of how to construct the whole evolving history of schemata given data stored in NoSQL databases, e.g., JSON stored in MongoDB. 
Instead of table schemata in relational databases, they consider the structure of persisted objects in NoSQL databases, referred to as \emph{entity types}.
The proposed approach first extracts each entity type from loaded  datasets, with assigned timestamps that indicate  its residing time interval. 
Then from different structure versions of the entity types, it detects the possible operations   between two consecutive versions. In the case of multiple alternative operations, users will make the final validation.
In addition, to detect certain schema changes, it is often useful to detect integrity constraints, e.g., inclusion dependencies. The assumption in \cite{DBLP:conf/bigdataconf/KlettkeAS0S17} is that in NoSQL databases often schemata are ``less'' normalized, 
  which leads to the inclusion dependencies involving multiple attributes rather than a single attribute as in relational databases. 
In \cite{DBLP:conf/bigdataconf/KlettkeAS0S17}  an algorithm is proposed to detect such k-ary inclusion dependencies.
\subsection{Data provenance  }
\label{ssec:provenance}
\emph{Data provenance} (also known as \emph{data lineage}) refers to meta information of data records, which indicates their origin, usage, status in the life cycle, etc. 
The provenance information can be seen as a special type of metadata, which tells how a dataset is obtained from  original sources and helps to make   proper access to  datasets \cite{simmhan2005survey}. 
Such information could be extracted during data ingestion, and later enriched during maintenance or exploration, possibly with human input.

In \cite{terrizzanodata}, a governance tool from IBM is presented, which can manage the requests for ingesting new data sources or using already ingested datasets in a data lake.  
Suriarachchi et al. \cite{DBLP:conf/eScience/SuriarachchiP16} propose an abstracted architecture that provides integrated provenance (information of activities) 
given multiple data processing and analytics systems (e.g., Hadoop, Storm\footnote{\url{https://storm.apache.org/}}, and Spark), 
as these systems populate provenance events in different standards and apply various storage manners. 
They also 
study a use case, in which  data from Twitter is collected and processed 
(e.g., count hashtags, aggregate data by each category) by Apache Flume\footnote{\url{https://flume.apache.org/}}, Hadoop jobs, and Spark jobs. 

GOODS \cite{DBLP:conf/sigmod/HalevyKNOPRW16, DBLP:journals/debu/HalevyKNOPRW16}, CoreDB \cite{DBLP:conf/cikm/BeheshtiBNCXZ17, DBLP:journals/pvldb/BeheshtiBNT18}, and Juneau \cite{DBLP:journals/pvldb/ZhangI19, DBLP:conf/cidr/IvesZHZ19, zhang2020finding} all preserve the provenance information as graphs. 
As mentioned in Sec.~\ref{sssec:catalog}, GOODS  stores provenance information in its metadata catalog, as one of the six metadata groups. It builds provenance graphs and visualizes them to users, such that users can keep track of the usage and transformation of the data. 
It  exports the provenance metadata in the catalog as subject–predicate–object triples into a graph-based system, then generates the provenance graphs for visualization and path-based  querying.  
CoreDB uses the descriptive, administrative and
temporal metadata to build   DAG-based provenance graphs \cite{beheshti2012temporal}, 
which helps answer questions such as who queried a specific entity.
%
Juneau \cite{zhang2020finding} generates graphs with variables  as nodes, and connects two variable nodes in the same function. %
Given a variable $v$ in the notebook, one can find all other variables affecting $v$ via some functions, and the relationships between these variables and $v$.

\section{Exploration}
\label{sec:exp}
It is important that   useful information can be retrieved from data lakes. 
However, this is often a challenging task due to the large number of ingested sources, and the heterogeneity of data. A user may have knowledge of one or a few data sources, but rarely, if not never, all the datasets. Thus, the existing solutions mainly solve the querying problem in data lakes in the following two directions: 
explore the data lakes based on the relatedness of datasets (Sec.~\ref{ssec:qds}), or provide a unified query interface for heterogeneous data sources (Sec.~\ref{ssec:qrwt}).

\subsection{Query-driven data discovery} 
\label{ssec:qds}
\emph{Query-driven data discovery} \cite{DBLP:journals/pvldb/Miller18} refers to searching a data lake based on the measured relatedness (e.g., joinable)  among datasets  as introduced in
Sec.~\ref{sssec:relDS}. 
With input queries specifying a given dataset (usually tabular data), the system returns the top-k most related datasets. 

\para{Exploration input/output} There are   three ways of exploration. We denote the set of datasets in a data lake as $\mathbf{S}$.
\begin{enumerate}[leftmargin=*]
	\item   Given the user-specified table $T$ and a column $c$ of $T$, the system returns top-k tables that are most related to $T$,  e.g., JOSIE \cite{DBLP:conf/sigmod/ZhuDNM19}.
	\item  Given 
	a table $T$, the system returns top-k tables (referred to as $\mathbf{S}^k$) that contain   relevant attributes for populating $T$, e.g., $D^3L$ \cite{bogatu2020dataset}. 
In addition, if a table $S_i$ is not in the top-k result set (i.e., $S_i \in (\mathbf{S}-\mathbf{S}^k)$),  yet it can be joined with some table(s) in $\mathbf{S}^k$ and improve the attribute coverage of $T$, 
$D^3L$ also  includes $S_i$ in the result. 
	\item   Given the user-specified table $T$ and the search type $\tau$ for  external applications (e.g., a data science task), the system returns top-k tables that are most relevant to $T$ based on the relatedness measurements associated to  $\tau$, e.g., \emph{Juneau} \cite{zhang2020finding}. 
\end{enumerate}
Notably, 
in this group of studies, the challenge of exploring datasets is  a ``search'' problem rather than a query reformulation problem in data integration.
It can also be seen as a  step prior to data integration or data science tasks \cite{bogatu2020dataset, zhang2020finding}.
 

\para{Querying methods and indexing}
Given an input query table, the systems in Table~\ref{tbl:dd} often rely on similarity estimation using indexes (e.g., aforementioned LSH indexes, inverted indexes). They  
rank candidate tables, and include the  top-k tables in the result.
In addition, Aurum \cite{DBLP:conf/icde/FernandezAKYMS18}  applies a graph index 
to accelerate expensive queries containing discovery path queries for searching its hypergraphs. In its primitive-based query language, an Aurum user can compose queries to search schemata or data values with  keywords to find specific columns, tables, or paths. Users can specify criteria and obtain ranked querying results in  a flexible manner, i.e., they can obtain the ranking results of different criteria without re-running the query. 
In  Juneau  \cite{zhang2020finding}, a query is  a cell output table picked by the user. 
The user also chooses the type of the search, e.g., find tables for data cleaning. 
Then Juneau uses the corresponding relatedness measurements to perform the top-k search. It speeds up the search with strategies such as \emph{indexing}  columns profiled in the same domain and tables connected by workflow steps, and \emph{pruning}  tables under a threshold of schema-level overlap.


\para{Remarks on future directions} Data lake exploration could benefit greatly by taking into account recent results on  web table exploration \cite{lehmberg2017stitching, cafarella2009data, pimplikar2012answering}, data wrangling \cite{terrizzanodata, furche2016data}, or external applications upon the data lake.
With the existing works mainly focusing on evaluating the accuracy of similarity computation, or the performance (query processing time), 
deep analysis and further improvement on the accuracy and completeness of the top-k result set are still rare. 
Finally, the existing solutions in this group mostly study tabular  data. 
In what follows, we discuss the data lakes that explore datasets with diverse data models.

\subsection{Heterogeneous data querying}
\label{ssec:qrwt}
In this survey, by \emph{querying heterogeneous data}, we indicate the systems providing a unified querying interface to access heterogeneously structured data.
Next, we introduce  studies that tackle such a research problem.

Constance \cite{hai2016constance, DBLP:conf/adbis/HaiQZ18} provides an incremental manner for users to explore the data lake. 
Via the user interface, a user can first browse the existing data sources, including their description, statistics, and schema; 
then she can write a query (SQL or JSONiq\footnote{\url{https://www.jsoniq.org/}}) for a single dataset. 
She can also  make a keyword search over the schemata  or the data. 
Alternatively, with certain knowledge of the datasets, which can be developed through the previous exploration processes, she can choose to integrate and query
a subset of datasets as introduced in Sec.~\ref{ssec:di}. 
In addition,    users can   transform  
data in the data lake to their desired structure and format. 
The  information retrieval requirements  of external applications are supported  via RESTful APIs \cite{DBLP:conf/dils/GeislerQHA17}. 

CoreDB \cite{DBLP:conf/cikm/BeheshtiBNCXZ17, DBLP:journals/pvldb/BeheshtiBNT18} 
provides users with a unified interface, i.e., through a REST API 
for querying data or performing  \emph{Create, Read, Update and Delete} (CRUD) operations. 
 It applies Elasticsearch\footnote{\url{https://www.elastic.co/}} 
for the underlying full-text search, SQL queries for relational database systems, and SPARQL queries for knowledge graphs.

\emph{Ontario} \cite{DBLP:conf/dexa/EndrisRVA19, handle:20.500.11811/8347} and \emph{Squerall} \cite{mami2019querying} both enable a federated query processing  over a semantic data lake and apply   SPARQL to query the  heterogeneous data lake.
Ontario supports heterogeneous data, e.g., RDF (stored in Virtuoso\footnote{\url{https://virtuoso.openlinksw.com/}}),   local JSON files,  
TSV files (in HDFS),  XML files (in  MySQL). 
It profiles each dataset  with its metadata and additional information, e.g., the types of the source, or the web API for querying this type of source. 
For instance, for TSV files stored in HDFS, it provides Spark-based services which translate the SPARQL queries to SQL. 
Given an input SPARQL query,  Ontario first decomposes the query. Then it uses the profiles to generate subqueries for each dataset with a set of proposed rules. Using metadata, it also tries to generate optimized query plans.
In \cite{rohde2020optimizing}, the general guidance of query optimization 
on top of Ontario is proposed.
Similar to Ontario, Squerall also supports querying diverse data sources, including files (i.e., CSV, Parquet) and relational (i.e., MySQL)  and NoSQL databases (i.e., Cassandra, MongoDB). The schemata of the sources are mapped to a mediator, which consists of high-level ontologies. 
Given SPARQL queries against the mediator, 
with the mapping the relevant data entities are retrieved from data sources, which are  joined and transformed to form the final query results.
Squerall enables distributed query processing and is implemented with two versions with different data connectors: Spark and Presto\footnote{\url{https://prestodb.io/}}.

\section{Challenges and future  directions } 
\label{sec:chall}
We have addressed specific technical aspects of managing data lakes so far. Next, we discuss the challenges of applying data lakes in \ broad technological application domains in Sec.~\ref{ssec:market}, and two main future research directions of data lakes in Sec.~\ref{ssec:DLML} and \ref{ssec:DM_more}.

\subsection{Data lakes in digital business transformation }
\label{ssec:market}
The organizational perspective is gaining relevance and places new requirements on future data lakes. Many large traditional firms, often incumbent market leaders in their (non-IT) business, pursue a digital transformation strategy in order to be able to compete with purely digital newcomers more  effectively. In this context, the classical model of using internal IT departments or external software houses via service contracts is increasingly abolished in favor of integrative product-oriented teams in which developers, domain experts, sales and purchase representatives closely work together in an agile manner within a longer-term business roadmap and architecture. Moreover, the digitization process involves frequent mergers, acquisitions, and re-organizations of the business. In many cases, traditional data and systems integration methods, together with different organizational philosophies, have led to the failure of such processes, in some cases costing billions of dollars.

Executives have begun to perceive the idea of data lakes as a design pattern to deal with this organizational volatility. In this pattern, the data lake (at its core a store of mildly cleaned raw data) serves as a mediating element between the evolving set of internal and external transaction and monitoring streams, and the equally evolving set of business analytics and decision support tasks of the above-mentioned teams. In the ideal case, integrating a new company would simply mean adding their raw data to the lake, and using methods discussed in our survey to link them up to the existing data lake content. Existing analytics solutions could automatically include the new data, and new business teams could use analytics toolkits plus specialist expertise for their current challenges, without waiting for resources in the IT department. In this way, executives are trying to convert the IT provisioning from a cost and reimbursement factor to a continuously value-creating investment \cite{ICIS22CIOPanel}.

Achieving such a setting can be mission-critical for many organizations in traditional businesses. However, its realization is by no means trivial, both on the organizational and on the technical side. The following  technical challenges concern both the input side and the analytics side of the data lake pattern.
%
First, nowadays, data is largely consumed by machine learning and data science applications everywhere. However, existing data lakes  lack  \emph{matured} functions to meet such a requirement. 
Second, the lack of traditional analytical data management such as transaction management, indexing, and caching,   makes data lakes less adequate for complex  analytical workloads in the industry. 
Tackling these limitations, we discuss the exciting new challenges next.


\subsection{Data lakes meet  machine learning}
\label{ssec:DLML}
Recent advances of \emph{DBML} \cite{Makrynioti2019survey, zhou2020database, 10.1007/978-3-030-35514-2_32}, i.e., in-database machine learning or applying machine learning for data management,  mainly consider a relational database instead of data lakes. 
Below we focus on the \emph{lake-specific} challenges. 

\para{Training data heterogeneity} The systems and methods covered in this survey mainly support  tabular data, JSON/XML, graphs, and texts.
However, ML training data may also include other common types such as images, audio, and videos. 
The challenge of data multi-modality is non-trivial, and stretches beyond simply utilizing technologies such as multimedia databases \cite{subrahmanian2012multimedia} and polystores.  
The key question is how to design abstractions for  heterogeneous data in data lakes. With the rapid development of ML models, e.g., BERT \cite{devlin2019bert}, GPT-3 \cite{brown2020language},  possible abstractions for multi-modal data are embeddings \cite{cui2018survey}. 
Such new  options  invoke more challenges. How to design the data abstraction to represent and connect multi-modal training data?  
How to design the data representation for a specific function in Table~\ref{tbl:all}, e.g., related dataset discovery, 
data integration?
Moreover, ML models
, in particular, deep neural networks require intensive tensor-based computations such as matrix multiplication.
With the recent advances in hardware, e.g., Tensor Processing Unit (TPU), and tensor runtimes such as ONNX\footnote{\url{https://onnxruntime.ai/}}, it is an interesting direction to explore tensor-based intermediate representations (IRs)  of data lakes, w.r.t. both data management and machine learning operations. It leads to more questions, e.g.,  how to redesign  data lake architectures based on such new intermediate representation possibilities?


\para{In-lake machine learning}
Following the research line of in-database machine learning \cite{schleich2019learning}, one of the most exciting challenges is to support machine learning training and inference in data lakes. First, the existing in-database machine learning studies mainly focus on structured data, i.e., relational tables. It is a rich area regarding how to extend these in-DB ML problems (e.g., factorization through joins \cite{10.1145/2882903.2882939, chen2017towards}) over heterogeneous, schema-less data in a data lake. Second, new APIs and systems are needed to connect existing data lakes with ML platforms such as MLflow\footnote{\url{https://mlflow.org/}}, Amazon SageMaker\footnote{\url{https://aws.amazon.com/sagemaker/}}, AzureML\footnote{\url{https://azure.microsoft.com/en-us/products/machine-learning}}, or \emph{model zoos} (repositories of pre-trained models) such as HuggingFace\footnote{\url{https://huggingface.co/}}, Tensorflow Hub\footnote{\url{https://www.tensorflow.org/}}, and PyTorch Hub\footnote{ \url{https://pytorch.org/hub/}}. 
A more ambitious design alternative is a tighter integration of data lakes and machine learning, i.e., \emph{ML-aware data lakes}, which are built upon the requirements of downstream ML applications, and bring more optimization opportunities. We need new data lakes providing functions such as preparing, labeling, and cleaning the raw,  heterogeneous data for downstream ML applications. To build ML-aware data lakes, it also calls for novel data lake architecture, storage, and function redesign. These considerations lead to the following research questions. How to discover related datasets to augment the existing training dataset and improve ML model accuracy and fairness? How to effectively clean the  raw,  heterogeneous datasets in data lakes  to improve the effectiveness of ML models? How to combine and optimize the whole pipeline of data management and  ML life cycle in data lakes?

\para{ML workflow optimization}
Towards designing ML-aware data lakes,  one of the main goals is to improve ML models in terms of effectiveness (e.g., model accuracy) and efficiency (e.g., training time). Besides the functional and system-level redesign, another possibility is to   utilize the   metadata. 
One of the most intensively studied DBML problems is optimizing   ML workflows and programs over  relational databases; surveys like  \cite{Makrynioti2019survey, zhou2020database} have elaborated on such studies.
For instance, by utilizing the primary key-foreign key relationships and join dependencies \cite{chen2017towards}, or functional dependencies  \cite{Khamis2020}, the runtime of   model training can be significantly reduced. However, one of the key differences  between data lakes from databases is the lack of metadata. 
Instead of predefined join dependencies or FDs,  in a data lake  we might need to discover the joinability  between datasets (Sec.\ref{ssec:early}) or relaxed functional dependencies (Sec.~\ref{ssec:rfd}), which are probabilistic. 
Thus, one of the open challenges of optimizing ML operations in data lakes, is to utilize such fuzzy, discovered metadata about data.

\para{ML-driven metadata management} 
Besides the metadata of data, we need to cover also the metadata  of ML models.
The life cycle of an ML model contains multiple steps, including model training, hyperparameter tuning, debugging, deployment, etc. 
Accordingly, we  need new metadata extraction, modeling, and enrichment methods for the relevant metadata about the ML life circle and the datasets involved in each step, which also calls for new data provenance methods. 


\subsection{Advanced analytics and transaction management}
\label{ssec:DM_more}

Another future direction is to bring the well-studied database and data warehouse functions, such as transaction management and query optimization, into data lakes for business intelligence. 
Towards this direction, the new paradigm of \emph{Lakehouse} \cite{armbrust2020delta, armbrust2021Lakehouse, jainanalyzing, hambardzumyandeep} has emerged. Earlier, a common industrial practice was to apply  data lakes (e.g., Amazon S3,  GCP) as a cheap storage of large-scale raw data, before the datasets are selected and transformed for data warehouses (e.g., Snowflake, BigQuery). The overhead and complexity of maintaining  two systems, a data lake and a data warehouse, have led to  Lakehouses, e.g., Delta lake \cite{armbrust2020delta}, Apache Hudi\footnote{\url{https://hudi.apache.org/}}, and Apache Iceberg\footnote{\url{https://iceberg.apache.org/}}. 
A Lakehouse inherits data lakes' role for storing large-scale raw data, i.e., supporting open formats such as Parquet and ORC over cloud storage, and data warehouses' analytics capabilities, e.g., transaction management, indexing, caching, and metadata management \cite{armbrust2020delta}. 


Following the path of Lakehouse, many challenges emerge regarding transaction management, storage, indexing, metadata management, and  machine learning.
How to design cloud-native storage for read-write workloads with low-latency transaction guarantees? 
How to design auxiliary structures such as indexes over open data formats for efficient query processing?
Moreover, recent Lakehouse \cite{jainanalyzing} provides open interfaces for ML workload to query the data, or tensor-based IR for deep learning models \cite{hambardzumyandeep}. A deeper integration between Lakehouse and ML, similar to the discussion in Sec.~\ref{ssec:DLML}, will bring more optimization opportunities, which calls for more research effort.



 \section{Conclusion and outlook} 
\label{sec:con}
In the first decade of their existence, 
data lakes have been receiving increasing interest from both  academia and industry.
In this survey, we have looked back at the origin and development of data lakes in the past decade. 
Besides offering a fine-grained data lake architecture and discussing storage systems,  we have provided  
a comprehensive review of existing  data lake methods based on their specific functions. 
We have used a three-level categorization, which facilitates  a deep analysis of the corresponding research questions. 
\new{To bring forth new challenges, we have also discussed potential technologies and future directions. 
} 

 \new{Without any doubt, the research, engineering, and application challenges  are there, waiting for novel data lakes to be developed together with   cutting-edge technologies of machine learning and cloud computing.} 
We foresee the explosive development of data lake  applications in the coming years.
 The golden age of data lakes is yet to come.


\IEEEdisplaynontitleabstractindextext

%
\IEEEpeerreviewmaketitle

\ifCLASSOPTIONcaptionsoff
  \newpage
\fi



%
\bibliographystyle{abbrv}
\bibliography{DLSurveyShorter}

%
\
\begin{IEEEbiography}[{\includegraphics[width=1in,height=1.25in,clip,keepaspectratio]{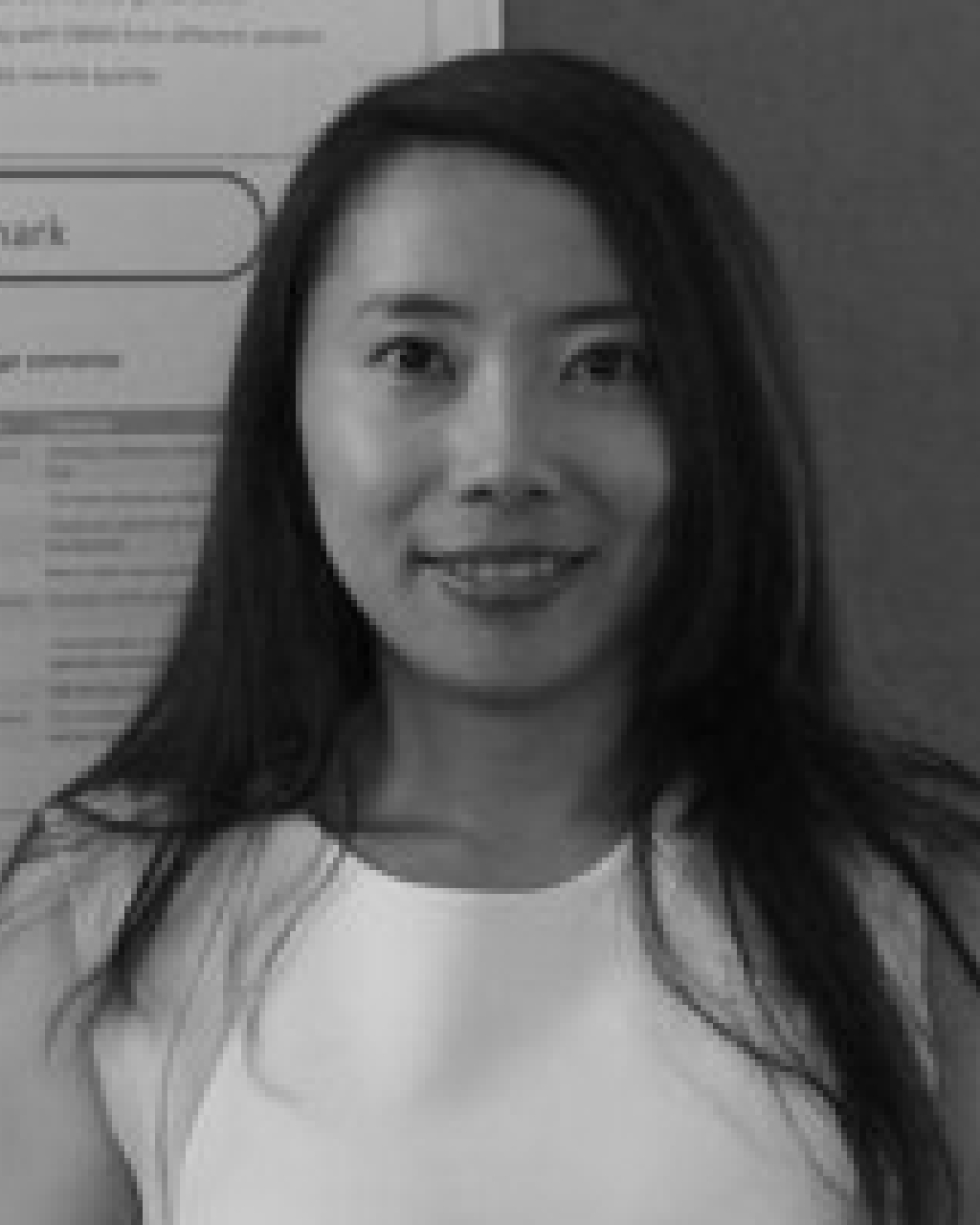}}]{Rihan Hai}
is an assistant professor in the Web Information Systems group at Delft University of Technology, Netherlands. She received her Ph.D. degree from  RWTH Aachen University, Germany. Her research focuses on data lakes, data integration, and related dataset discovery.  She has served as a program committee member of database conferences such as  ICDE and EDBT, and a journal reviewer for TKDE,  SIGMOD Record, JMLR and TPDS.
\end{IEEEbiography}
\vspace{-0.4cm}	
\begin{IEEEbiography}[{\includegraphics[width=1in,height=1.25in,clip,keepaspectratio]{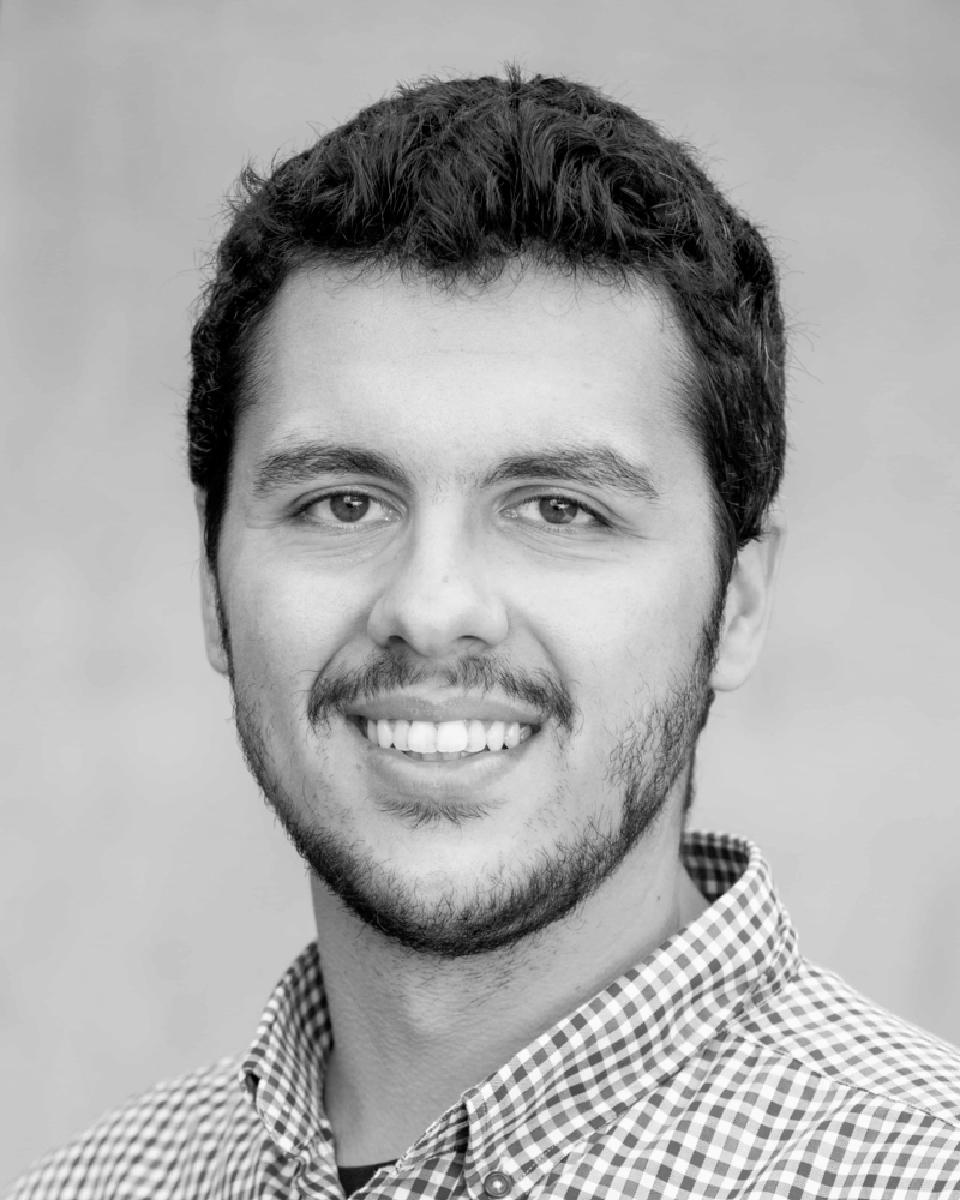}}]{Christos Koutras} received the five-year diploma in Electrical and Computer Engineering from NTUA, Greece, and the MPhil degree from the Department of Computer Science and Engineering, HKUST, Hong Kong. He is currently a PhD candidate in the Department of Software Technology at Delft University of Technology, Netherlands. His research interests include data integration, schema matching and related dataset discovery.
\end{IEEEbiography}
\vspace{-0.4cm}	
\begin{IEEEbiography}[{\includegraphics[width=1in,height=1.25in,clip,keepaspectratio]{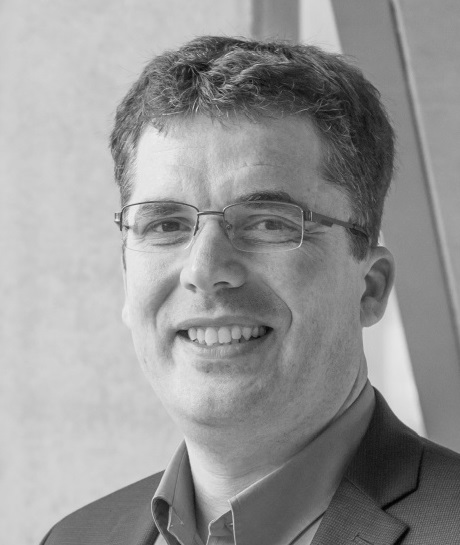}}]{Christoph Quix}
Christoph Quix is Professor for Information Systems and Data Science at the Niederrhein University.
Previously, he held a deputy professorship for data science at RWTH Aachen University.
His research focuses on data integration, data science, management of large, heterogeneous data sets, and metadata management. His research is application-oriented, e.g., in chemistry 4.0 or industry 4.0. He has more than 100 publications in international   journals and conferences. 
\end{IEEEbiography}
\vspace{-0.4cm}	

\begin{IEEEbiography}[{\includegraphics[width=1in,height=1.25in,clip,keepaspectratio]{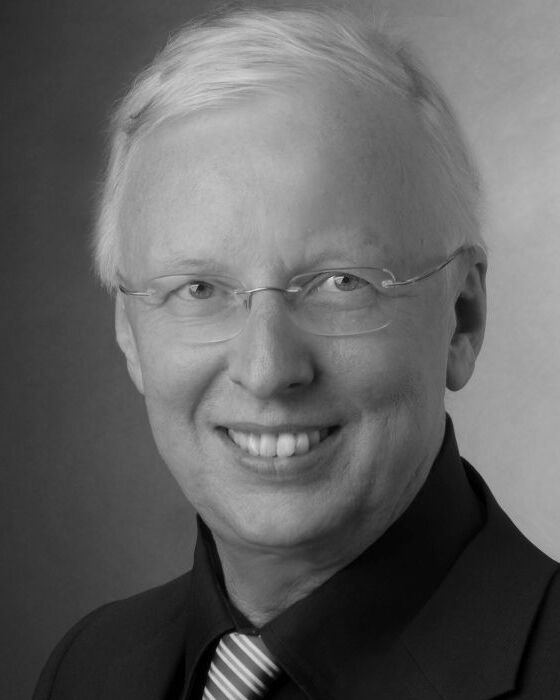}}]{Matthias Jarke}
Matthias Jarke is Professor Emeritus of Databases and Information Systems at RWTH Aachen University, Germany, and Past Director of the Fraunhofer FIT Institute for Applied IT. 
His research is focused on information systems support for cooperative tasks in engineering, business, and culture. Major contributions include query optimization, conceptual modeling, and requirements management. 
 Matthias has served as chief editor of Information Systems, and on the editorial board of numerous journals including IEEE TSE, as well as program chair of prestigious database conferences such as VLDB, EDBT, CAiSE, and SSDBM. 
He is lifetime senior IEEE member, ACM Fellow, GI Fellow, recipient of the 2020 Peter Chen Award, and member of Germany’s acatech National Academy of Science and Engineering.
\end{IEEEbiography}




\end{document}